\documentclass[12pt,preprint]{aastex}

\shorttitle{L1448C protostellar jet}
\shortauthors{Hirano et al.}

\begin{document}

\title{Extreme active molecular jets in L\,1448C}

\author{NAOMI HIRANO\altaffilmark{1,2}, PAUL P.T. HO\altaffilmark{1,3}, SHENG-YUAN LIU\altaffilmark{1}, HSIEN SHANG\altaffilmark{1},
CHIN-FEI LEE\altaffilmark{1}, 
\& TYLER L. BOURKE\altaffilmark{3}}

\altaffiltext{1}{Academia Sinica,
Institute of Astronomy \& Astrophysics, P.O. Box 23--141, Taipei, 106, 
   Taiwan, R.O.C.}
\altaffiltext{2}{e-mail: hirano@asiaa.sinica.edu.tw}

\altaffiltext{3}{Harvard-Smithsonian Center for Astrophysics, 60 Garden
   Street, Cambridge, MA 02138, USA}

\begin{abstract}
The protostellar jet driven by L1448C was observed in the SiO $J$=8--7 and CO $J$=3--2 lines and 350 GHz dust continuum at $\sim$1\arcsec resolution with the Submillimeter Array (SMA). 
A narrow jet from the northern source L1448C(N) was observed in the SiO and the high-velocity CO. 
The jet consists of a chain of emission knots with an inter-knot spacing of $\sim$2\arcsec (500 AU) and a semi-periodic velocity variation.  
These knots are likely to be the internal bow shocks in the jet beam that were formed due to the periodic variation of the ejection velocity with a period of $\sim$15---20 yr. 
The innermost pair of knots, which are significant in the SiO map but barely seen in the CO, are located at $\sim$1\arcsec (250 AU) from the central source, L1448C(N). 
Since the dynamical time scale for the innermost pair is only $\sim$10 yr, SiO may have been formed in the protostellar wind through the gas-phase reaction, or been formed on the dust grain and directly released into the gas phase by means of shocks. 
It is found that the jet is extremely active with a mechanical luminosity of $\sim$7 $L_{\odot}$, which is comparable to the bolometric luminosity of the central source (7.5 $L_{\odot}$). 
The mass accretion rate onto the protostar derived from the mass-loss rate is $\sim$10$^{-5}$ $M_{\odot}$ yr$^{-1}$.
Such a high mass accretion rate suggests that the mass and the age of the central star are 0.03--0.09 $M_{\odot}$ and (4--12)$\times$10$^3$ yr, respectively, implying that the central star is in the very early stage of protostellar evolution.
The low-velocity CO emission delineates two V-shaped shells with a common apex at L1448C(N).
The kinematics of these shells are reproduced by the model of a wide opening angle wind.
The co-existence of the highly-collimated jets and the wide-opening angle shells can be explained by the ^^ ^^ unified {\it X}-wind model'' in which highly-collimated jet components correspond to the on-axis density enhancement of the wide-opening angle wind.
The CO $J$=3--2 map also revealed the second outflow driven by the southern source L1448C(S) located at $\sim$8.3\arcsec (2000 AU) from L1448C(N). 
Although L1448C(S) is brighter than L1448C(N) in the mid-IR bands, the momentum flux of the outflow from L1448C(S) is two or three orders of magnitude smaller than that of the L1448C(N) outflow. 
It is likely that the evolution of L1448C(S) has been strongly affected by the powerful outflow from L1448C(N).%
\end{abstract}


\section{INTRODUCTION}
It is known that several deeply-embedded young stellar objects are driving highly-collimated molecular outflows with an extremely-high velocity (EHV) and jet-like component running along the axes of the lobes \citep[e.g.,][]{Bac96}.  
Since the EHV molecular jets having the terminal velocities of 50--150 km s$^{-1}$ are concentrated within the narrow angles along the axes of the lobes and have large momenta comparable to those of the slowly moving (20--30 km s$^{-1}$) ^^ ^^ classical'' outflows, the EHV jets are considered to be closely connected to the ^^ ^^ primary jet'' which is responsible for driving molecular outflows.
Therefore, studying the physical and kinematical properties of the EHV jet will allow us to understand the properties of the ^^ ^^ primary jet'' which provide clues to constrain the launching mechanism of the outflows.

The low-luminosity (7.5 $L_{\odot}$; \citet{Tob07}) class 0 source L1448C (also known as L1448-mm) in the Perseus molecular cloud complex (D$\sim$250 pc: e.g., \citet{Eno06}) is a spectacular example of an outflow with an EHV jet.
The EHV component of this source was identified as the secondary peaks of the CO $J$=2--1 spectra at ${\sim}{\pm}$60 km s$^{-1}$ from the cloud systemic velocity \citep{Bac90}.
The CO emission in the EHV range was found to be confined to a series of discrete clumps, called ^^ ^^ bullets", which are aligned along the axes of the outflow lobes and are symmetrically placed with respect to the central source.
The EHV bullets are also observed in several transitions of the SiO \citep[e.g.][]{Bac91, Dut97}.
Since the SiO emission has been barely detected in quiescent dark clouds because of its very low abundance (of the order of 10$^{-12}$; \citet{Ziu89,Mar92}), the detection of SiO in EHV bullets suggests the presence of shocks that enhanced the SiO abundance in bullet gas by a factor of $\gtrsim$10$^4$.
Although the lower transition of SiO, i.e. $J$=2--1,  was observed not only in the EHV bullets but also in the lower velocity component that delineates the tips and walls of the outflow cavities, $J$=5--4 emission was confined to a pair of EHV bullets located at the closest positions to the star \citep{Bac91}. 
This suggests that the excitation condition of the EHV jet varies along the axes, and that the jet gas is highly excited in the close vicinity of the driving source.
Recently, higher excitation SiO up to $J$=11-10 has been observed by \citet{Nis07}.
Their results have revealed that the innermost pair of bullets, labeled B1 and R1 by \citet{Bac90}, have a density of $n_{\rm H_2}\sim10^6$ cm$^{-3}$ and a kinetic temperature of $T_{\rm kin}\gtrsim$ 500 K, which is denser and warmer than the bullets in the downstream.
It is also known that the innermost pair of bullets, B1 and R1 are resolved into two clumps, BI-BII and RI-RII, respectively, in the higher resolution ($\sim$2\arcsec) interferometric SiO $J$=2--1 observations \citep{Gui92, Gir01}.
The high resolution SiO data exhibit the kinematic structure of the EHV jet near the source, which shows an apparent acceleration of the jet up to 70 km s$^{-1}$ within a region of 6$''$ ($\sim$2000 AU).
The proper motion measurement of the SiO clumps carried out by \citet{Gir01} suggests that the outflow axis is inclined by 21$^{\circ}$ with respect to the plane of the sky, and therefore, the SiO clumps in the EHV jet are likely to be moving with absolute velocities of 180 km s$^{-1}$.

In this paper, we present the SiO $J$=8--7, CO $J$=3--2, and 350 GHz continuum images obtained with the Submillimeter Array\footnote{The Submillimeter Array is a joint project between the Smithsonian Astrophysical Observatory and the Academia Sinica Institute of Astronomy and Astrophysics and is funded by the Smithsonian Institution and the Academia Sinica.} (SMA) \citep{Ho04} at $\sim$1 arcsecond resolution, which is a factor of three higher than the previous SiO $J$=2--1 images \citep{Gui92, Gir01}.
A high angular resolution is crucial for studying the structure and kinematics of the jet near the base, at which the jet velocity increases up to the terminal velocity.
In addition, higher transitions of SiO and CO in the submillimeter waverange enable us to segregate the dense and warm gas in the EHV jet from the lower excitation gas in the cavity wall.

\section{OBSERVATIONS}

The observations of the SiO $J$=8--7 and CO $J$=3--2 lines and 350 GHz continuum emission were carried out with the SMA on 2006 December 5 in the extended configuration and on 2006 December 25 in the compact configuration.
The two array configurations provided projected baselines ranging from 12\,m to 222\,m.
Since the primary beam of the SMA antenna has a size of $\sim$35\arcsec, two pointings separated by 17\arcsec  were observed in order to cover the EHV bullet pair closest to the central source.
The receivers have two sidebands; the lower and upper sidebands covered the frequency ranges from 345.5 to 347.5 GHz, and from 355.5 to 357.5 GHz, respectively.
The SiO $J$=8--7 and CO $J$=3--2 lines were observed simultaneously in the lower sideband.
The SMA correlator divides each sideband of 2 GHz bandwidth into 24 ^^ ^^ chunks" of 104 MHz width.
We used the configuration that gave 256 channels to all chunks, which provided a uniform frequency resolution of 406.25\,kHz across a 2\,GHz-wide band.
The corresponding velocity resolution was 0.35 km s$^{-1}$.
We used Titan for flux calibration, and a pair of quasars 3C84 and 3C111 for amplitude and phase calibrations.
The flux calibration was estimated to be accurate to 25\%.
The band pass was calibrated by observing 3C273.

The calibrated visibility data were Fourier transformed and CLEANed using the MIRIAD package.
The velocity-channel maps of the SiO and CO were made with a velocity interval of 1 km s$^{-1}$.
The synthesized beam size of the SiO map was 0\farcs96$\times$0\farcs84 with a position angle of $-$84$^{\circ}$ and that of the CO map was 0\farcs96$\times$0\farcs86 with a position angle of $-$76$^{\circ}$ with uniform weighting.
The rms noise level of the velocity-channel map at 1.0 km s$^{-1}$ width was 0.15 Jy beam$^{-1}$.
A non-linear joint deconvolution, MOSSDI, which is based on the CLEAN-based algorithm, and is part of the MIRIAD package \citep{Sau96}, was used for deconvolving the images. 

The 350 GHz continuum map was obtained by averaging the line-free chunks of both sidebands.
To improve the signal to noise ratio, the upper and lower sidebands data were combined.
The synthesized beam size of the map made with uniform weighting was 0\farcs93$\times$0\farcs83 with a position angle of $-$86$^{\circ}$.
The rms noise levels of the 345 GHz continuum maps was 6.4 mJy beam$^{-1}$.

\section{RESULTS}
\subsection{Continuum emission}

The 350 GHz continuum map reveals a bright compact source at the center.
This source has a peak intensity of 352 mJy beam$^{-1}$, and is surrounded by spatially extended emission.
In addition to this bright source, there is a faint emission peak of $\sim$5$\sigma$ level at $\sim$8\farcs3 southeast of the center.
Recent {\it Spitzer Space Telescope} observations at mid-infrared wavelengths have resolved L1448C into two components, L1448C(N) and L1448C(S) \citep{Jor06}.
The bright submillimeter source at the center corresponds to L1448C(N), which is considered to be the driving source of the highly-collimated molecular outflow and is also referred to as L1448-mm, and the southern faint source corresponds to L1448C(S).
Dust continuum emission from L1448C(S) was also detected by \citet{Jor07} at 230 GHz and 350 GHz with the SMA, and by \citet{Mau10} at 107 GHz with the Plateau de Bure Interferometer (PdBI).

The visibility amplitudes plot for L1448C(N) as a function of {\it uv} distance (Fig.~\ref{fig2}) suggests that this source consists of two components; one is from a spatially extended envelope that dominates the flux at a {\it uv} distance of $<$50 k$\lambda$, and the other is from a compact source that is prominent at $>$50 k$\lambda$.
The visibility amplitude profile was fit by two circular gaussian components; one is an extended component with a deconvolved size of $\sim$3.6\arcsec and a flux of $\sim$440 mJy, and the other is a compact component with a deconvolved size of $\sim$0.3\arcsec and a flux of $\sim$330 mJy.
The peak flux values of the extended and compact components correspond to $\sim$27 mJy and $\sim$330 mJy beam$^{-1}$, respectively, per 0\farcs93$\times$0\farcs83 beam.
As shown in Fig. 1a, the extended component appears as a bump in the northwest and a tail in the southeast (both of them are in the 3$\sigma$ level) of the central source, suggesting that this emission extends along the outflow axis.
Similar emission feature along the outflow axis is also seen in the maps of 3 mm \citep{Gui92}, 2.6 mm \citep{Bac95}, 1.4 mm \citep{Sch04}, and 1.3 mm \citep{Mau10} observed with 1.5--3\arcsec resolution.
Such a faint elongated feature was clearly seen in L1157-mm, and was interpreted as the edges of the cavity excavated by the outflow \citep{Gue97}.
It is, therefore, possible that the extended emission in L1448C(N) also delineates the inner part of the envelope which is disturbed by the outflow, although the cavity-like structure is not clearly seen.
An alternative interpretation is that the faint emission comes from an embedded companion.
A faint secondary peak seen in the 230 GHz map of \citet{Jor07} implies this possibility.
However, the position of the secondary peak at 230 GHz in \citet{Jor07} is 1\arcsec offset toward the northwest from that of our 350 GHz map.
As discussed in \citet{Jor07} such a small difference in position between the 350 GHz peak and 230 GHz peak could be introduced by the extended emission from the envelope component sampled by different {\it uv} coverages.
Since the total flux at 350 GHz observed with the SMA (extended + compact components) corresponds to 43 \% of the flux observed by \citet{Hat05} at 850 $\mu$m with SCUBA at the JCMT (1.737 Jy per 14\arcsec  beam), it is possible that the extended component that is not sampled well with the SMA affects the morphology of the faint component.
Although the northwestern bump may harbor an embedded companion, it is likely that most of the extended emission arises from the inner part of the envelope.
Therefore, the former scenario, which attributes the extended component to the envelope--outflow interacting region, is more preferable in this case.

The visibility amplitude profile at {\it uv} distance longer than $\sim$50 k$\lambda$ implies that the compact component is not point-like but has a spatially resolved structure.
In order to exclude the contamination from the extended envelope component, we made a map using the visibility data with the {\it uv} distances greater than 70 k$\lambda$ (see Fig.1b).
The synthesized beam of this map is 0\farcs70$\times$0\farcs50 with a position angle of $-$87.2$^{\circ}$.
It is shown that the compact source has an elongation along the axis perpendicular to the outflow axis.
A two dimensional Gaussian fit to the visibility data with {\it uv} distances larger than 70 k$\lambda$ yields the deconvolved major and minor axes of 0.37\arcsec (90 AU) and 0.26\arcsec (65 AU), respectively, with a position angle of $\sim$70$^{\circ}$.
The source position derived from the fit is $\alpha$(J2000) = 3$^h$25$^m$38.873$^s$, $\delta$(J2000)=30$^{\circ}$44$'$05\farcs35.
This position agrees well (less than 0\farcs1) with the 3.6 cm continuum position observed with a smaller beam of 0\farcs31$\times$0\farcs27 \citep{Rei02}.

The 350 GHz flux of L1448C(S) was measured to be $\sim$60 mJy, which is consistent to that reported by \citet{Jor07}.
L1448C(S) was detected in 230 GHz by \citet{Jor07} but not by \citet{Mau10} at the same frequency.
The non-detection of this source by \citet{Mau10} is probably because this source is located near the edge of the primary beam of the PdBI ($\sim$22\arcsec).
If the response of their primary beam is taken into account, the 230 GHz continuum source with a flux of 12.8$\pm$3 mJy detected by \citet{Jor07} could be below the detection limit of their observations ($\sim$8.4 mJy beam$^{-1}$).
Although L1448C(S) is bright in mid-infrared, its sub-mm flux is more than ten times weaker than L1448C(N).
It is unlikely that such a small sub-mm flux is due to the effect of missing flux, because single-dish measurements at 450 and 350 $\mu$m \citep{Cha00} showed no hint of the secondary component to the south of L1448C(N) even though the angular resolution ($\sim$8\arcsec) was comparable to the separation of two sources.

\subsection{The SiO Jet}

The SiO $J$=8--7 emission was detected in two velocity ranges from $-$70 to $-$12 km s$^{-1}$ (blueshifted) and from 20 to 71 km s$^{-1}$ (redshifted) with respect to the systemic velocity of $V_{\rm LSR}{\sim}$5.0 km s$^{-1}$.
Fig.~\ref{fig3} shows velocity channel maps of the SiO $J$=8--7 emission at 10 km s$^{-1}$ intervals.
It is shown that the SiO emission comes from the jetlike narrow region with its blueshifted part to the northwest and the redshifted part to the southeast of L1448C(N).
The SiO $J$=8--7 jet is partially resolved along its minor axis; after deconvolution from the SMA beam, the width of the SiO jet is $\sim$0.8\arcsec ($\sim$200 AU) FWHM on average.
In order to estimate the missing flux, we have smoothed the SMA map to a resolution of 14\arcsec and compared it with the SiO $J$=8--7 spectra observed with the James Clerk Maxwell Telescope \citep[JCMT;][]{Nis07}.
It is found that 85--100 \% of the single-dish flux is recovered by the SMA, implying that almost all the SiO $J$=8--7 emission arises from the narrow jet.
SiO $J$=8--7 emission mainly comes from the B1 and R1 ^^ ^^ bullets'' identified in the single-dish CO $J$=2--1 map by \citet{Bac90}.
The SiO jet consists of a chain of knots with a typical size scale of $\sim$1--1.5\arcsec.
A comparison with the previous SiO $J$=2--1 maps with $\sim$3\arcsec resolution \citep{Gui92, Gir01} reveals that the three pairs of knots close to L1448C(N) seen in the SiO $J$=8--7 map corresponds to the inner pair BI and RI in the SiO $J$=2--1 map, and the two pairs in the downstream correspond to the outer pair BII and RII.
The innermost knot pair BIa and RIa are located within 1\arcsec (250 AU) from L1448C(N).
The high-resolution image also shows that the SiO jet is not straight.
A close-up view of the high velocity component (Fig.~\ref{fig5}) shows that the jet changes its position angle from +15$^{\circ}$ at BII, $-$25$^{\circ}$ at BI, $-$20$^{\circ}$at RI, to $-$5$^{\circ}$ at RII.
The kinks between BI and BII, and RI and RII are also seen in the previous SiO $J$=2--1 maps of $\sim$3\arcsec resolution \citep{Gui92, Gir01}.
However, it is more obvious in the higher resolution image.
In addition, it is clear that the jet axes in BI and RI are also misaligned by 5$^{\circ}$.
\subsection{CO Jet and outflow}

The CO $J$=3--2 emission was detected in the wide velocity range from $-$77 km s$^{-1}$ to $+$79 km s$^{-1}$ with respect to the systemic velocity.
In the velocity ranges of ${\Delta}V < {\pm}$40 km s$^{-1}$ (Fig. 4a--4d), the CO $J$=3--2 emission delineates two V-shaped structures  open to the northwest and southeast with a common apex at the position of L1448C(N).
The opening angles of the V-shape features become narrower as the velocity offset increases, suggesting that the CO emission in the V-shaped features comes from the limb-brightened shells.
The largest opening angle of the shell is $\sim$60$^{\circ}$ in the blueshifted lobe, while it is $\sim$40$^{\circ}$ in the redshifted lobe.
In the higher velocity ranges of ${\Delta}V = {\pm}$41--70 km s$^{-1}$ (Fig. 4e, 4f, and 4g), the CO emission comes from a narrow jet-like region.
The CO flux recovered by the SMA in each velocity range was estimate by comparing the CO spectra observed by the SMA with those observed by the JCMT.
The SMA map was smoothed to be a 14\arcsec resolution so as to match with the beam of the JCMT.
It is found that the recovered flux is only $\sim$20~\% in the lowest velocity ranges (${\Delta}V = {\pm}$1--10 km s$^{-1}$; Fig. 4a), and is $\sim$50\% in the next velocity range of ${\Delta}V = {\pm}$11--20 km s$^{-1}$ (Fig. 4b). 
The edges of the shells in Fig. 4a and 4b look very steep probably because significant amount of the CO emission from spatially extended component was filtered out by the interferometer.
In fact, previous examples of the L1157 outflow and IRAS 04166+2706 outflow revealed that the edges of the bipolar cavities were emphasized in the maps made with the interferometer data alone, while the cavities were filled by the diffuse CO emission when the single-dish data were added \citep{Gue96,San09}.
On the other hand, in the higher velocity ranges with ${\Delta}V > {\pm}$20 km s$^{-1}$, 80--100\% of the CO flux was recovered by the SMA.
This suggests that almost all the CO $J$=3--2 flux in the extremely high velocity ranges (Fig. 4e, 4f, and 4g) comes from the narrow jet.

As in the case of the SiO jet, the CO $J$=3--2 jet also shows clumpy structure.
In addition, most of the knots seen in the SiO $J$=8--7 map have their counterparts in the CO map.
However, the innermost knots BIa and RIa, which are significant in the SiO map at ${\Delta}V = {\pm}$21--60 km s$^{-1}$ are barely seen in the CO map.
This is probably because most of the CO molecules in these knots are excited to the levels higher than $J$=3 because of high density and high temperature.
Similar feature with strong SiO and weak CO in the close vicinity of the protostar was also observed in the highly collimated jet in the HH211 outflow \citep{Pal06, Lee07b}.
Another difference between the CO jet and SiO jet is seen between BIc and BIIa, and RIc and RIIa. 
The CO map reveals the knot pair labeled BI-II and RI-II, while the SiO map shows only faint emission (Fig.~\ref{fig5}).
In the CO $J$=3--2, the overall distribution of the EHV jet is more continuous as compared to the SiO distribution. 

As in the case of the SiO jet, the blue and red axes of the CO jet are also misaligned by $\sim$5$^{\circ}$.
The kinks between BI and BII, and RI and RII are also seen in the CO jet.
In addition, the ridge of the CO emission wiggles along the jet axis.
This wiggling feature is clearly seen in the maps of the highest velocity ranges (Fig.~\ref{fig6}).
It is likely that each knot has been ejected in slightly different direction.
Since the typical knot separation, 2--3\arcsec corresponds to a time interval of 15--20 yr, the observed jet wiggling suggests that the direction of jet ejection also varies in a similar time scale.
On the other hand, the CO emission with lower intensity that is surrounding the emission ridge tends to extend linearly along the axes.
The transverse width of the lower level CO emission component increases with a distance from the source.
This is probably because significant part of the CO emission comes from the outflow shell even in the extremely high velocity ranges (see the next section).
There is faint CO emission with the highest velocity seen ahead of the SiO jet.
Since this highest velocity emission is spatially extended, it is likely that it arises from the highest velocity part of the shells.

\subsection{Kinematics of the jet along its axis}

Position-velocity (P-V) diagrams of the SiO and CO emission along the jet axes (the position angle is $-$25$^{\circ}$ for the blueshifted part and is $-$20$^{\circ}$ for the redshifted part) are shown in Fig.~\ref{fig7}.
Due to the change of the position angle, the outer knots BIIb and RIIb do not appear in these P-V diagrams.

The velocity structure of the jet is well traced in the SiO.
The jet velocity rapidly increases within $\backsimeq$1\arcsec from the star, and reaches close to its highest velocity (${\pm}$65 km s$^{-1}$ from $V_{\rm sys}$) at $\sim$5\arcsec from the star, which corresponds to the positions of the BIc and RIc knots.
The velocity dispersion is extremely large (${\Delta}V{\sim}$50 km s$^{-1}$ at 1$\sigma$ level) at the base, while it narrows to ${\Delta}V{\sim}$20 km s$^{-1}$ at the positions of the BIc and RIc knots. 
In the positions further downsteam, from BIc to BIIa and from RIc to RIIa, the observed radial velocity slightly decreases in the blueshifted side, while it increases in the redshifted side.
These radial velocity changes are probably due to the change of inclination angle, because the position angle of the jet also changes at these positions.
The velocity pattern shown in the SiO is similar to those of the atomic jets from class I source and T Tauri stars, in which the low-velocity components with broad line widths are located near the base and the high-velocity components with narrower line widths are located further from the source \citep[e.g.][]{Pyo02}.
In addition to the global velocity structure, each knot shows its internal velocity gradient with its higher velocity in the upstream side and lower velocity in the downstream side.

\subsection{Kinematics of the CO outflow shells}

In the CO $J$=3--2, the jet shows a similar velocity pattern with a similar velocity centroid as the SiO.
In addition to the jet, the P-V map of the CO shows extended low velocity features (slower than $\pm$15 km s$^{-1}$) and linear velocity features with the velocity magnitude increasing with the distance from the source (^^ ^^ Hubble-law''). 
These Hubble-law features are obviously different from the jets, and are not seen in the SiO, suggesting that they are from the outflow shells.
It should be noted that the radial velocity offsets of the linear velocity components become larger than 50 km s$^{-1}$ at around $\backsimeq$10\arcsec from the star, and contaminate the CO map in the EHV range.

The observed velocity pattern of the outflow shell is different from that predicted by the jet-driven bow shock model, in which the P-V structure shows a convex spur with high-velocity components at the jet head \citep[e.g.][]{Mas93, Che94}; it is rather similar to the velocity pattern produced by the wide-angle wind model \citep{Shu91, Shu00, Li96}.
In the wide-angle wind model, an outflow shell consists of ambient material swept-up by a radial wind from a young star.
A ^^ ^^ Hubble-law'' in the shell velocity is expected if the shell is expanding into an ambient medium with a radial density profile of ${\propto}r^{-2}$ \citep{Shu91}.
The shape of the shell is determined by the combination of the poloidal density profiles of the wind and ambient gas.
It is approximately parabolic for an $X$-wind type of wide opening angle wind with an angle-dependent density profile of ${\propto}1/{\rm sin}^2{\theta}$ (where $\theta$ is an angle measured from the axis of the flow) expanding into an ambient medium with a ${\propto}{\rm sin}^2{\theta}/r^2$ density profile, which is appropriate for magnetized cores \citep{Li96}.
Therefore, we adopted the simplified analytical model of a wind-driven model proposed by \citet{Lee00} to examine whether the observed morphology and kinematics of the CO outflow shells can be explained by means of wide-opening angle wind model.
In the cylindrical coordinate system, the structure and velocity of the shell can be written as follows:
 \begin{equation} z=CR^2, {\it v}_R={\it
v}_0R, {\it v}_z={\it v}_0z, \end{equation} 
where z is the distance along the outflow axis; R is the radial size of the outflow perpendicular to z ;
$C$ and {\it v}$_0$ are free parameters which describe the spatial and velocity distributions of the outflow shell, respectively.
The observed outflow shell features and the velocity patterns of the redshifted and blueshifted lobes were successfully reproduced by the model curves with $C$=0.8 arcsec$^{-1}$and {\it v}$_0$=5.0 km s$^{-1}$ arcsec$^{-1}$, and $C$=0.6 arcsec$^{-1}$ and {\it v}$_0$=5.0 km s$^{-1}$ arcsec$^{-1}$, respectively.
Here the inclination angle of the outflow axis with respect to the plane of the sky was assumed to be 21$^{\circ}$, which is derived from the proper motion measurement for the SiO $J$=2--1 jet by \citet{Gir01}.
The dynamical age of the outflow shell is given by 1/$v_0$ and is estimated to be $\sim$240 yr.
This number is roughly consistent with the dynamical age of $\sim$500 yr derived from the extent of the shell ($\sim$25\arcsec = 6300 AU), the mass-weighted-mean radial velocity of the shell ($\sim$22 km s$^{-1}$), and the inclination angle of the outflow axis ($\sim$21$^{\circ}$).
The model curves that delineate the outer boundaries of the lobes projected onto the plane of the sky are shown in Fig.~\ref{fig8} on top of the contours of the outflow shells.
The model curves of the P-V maps are shown in Fig.~\ref{fig7}.
However, this simplified wind-driven model has difficulty in reproducing several observed features.
First, the observed CO intensity drops sharply at the systemic velocity and does not extend to the opposite velocity ranges, while the model curves on the P-V maps predict the emission at $V_{\rm LSR}{\sim}$10 km s$^{-1}$ in the blueshifted lobe and at $V_{\rm LSR}{\sim}$0 km s$^{-1}$ in the redshifted lobe.
Second, the shapes of the shells in the different velocities cannot be reproduced (Fig.~\ref{fig9}).
The observed CO emission shows V-shaped distributions in the velocity ranges close to the systemic velocity.
On the other hand, the model curves predict elliptical shapes (Fig. 9a and 9b).
In addition, the transverse width of the CO shell becomes narrower as the velocity offset increases, while the model predict the opposite trend.
It should be noted that the CO emission near the protostar in Fig. 9c and 9d arises from the jet.

\subsection{Kinematics across the jet}

In order to search for the signs of jet rotation, the velocity gradient along the minor axes of the jet was examined.
The P-V diagrams of the SiO and CO at the positions of innermost two pairs of knots (Fig.~\ref{fig11}) show no clear velocity gradient at the positions of BIa, RIa, and RIb.
On the other hand, there is some hint of velocity gradient in the SiO at the position of BIb;  the southwestern side of the jet tends to be more blueshifted than the northeastern part.
However, this velocity gradient is in the opposite sense to the rotation pattern of the NH$_3$ core with its blueshifted part in the northeast and the redshifted part in the southwest \citep{Cur99}.
This suggests that the observed velocity gradient is not due to the rotation.

\subsection{Physical properties of the EHV jet}

The physical parameters of the EHV jet were estimated using the CO flux measured in the velocity ranges of ${\Delta}V$=50--70 km s$^{-1}$, in which most of the CO flux was recovered by the SMA.
We assumed that the CO emission in these velocity ranges is optically thin, and that the excitation condition of the CO follows the LTE.
A fractional abundance of the CO and a mean atomic weight were adopted to be 10$^{-4}$ and 1.41, respectively.
The kinetic temperature of molecular gas in the EHV jetwas derived to be 500--1200 K from the far infrared lines of CO, H$_2$O, and H$_2$ \citep{Nis99,Nis00}, and $>$500 K from the millimeter and submillimeter SiO lines \citep{Nis07}.
However, it is uncertain whether the CO (3-2) emission arises from the same gas component that contributes to the higher transition lines in the far-infrared.
In fact, \citet{Gus08a} have modeled the multi-transition SiO data observed by \citet{Nis07} using their face-on C-type shock models, and obtained much lower temperature of 70--90 K.
This implies that the most of the gas in the EHV jet is not so warm as $>$500 K. 
Therefore, an excitation temperature of $\sim$100 K was assumed for the calculation.

The mass for each of the blueshifted and the redshifted jets is estimated to be $\sim$10$^{-3}$ $M_{\odot}$, which is a few times higher than the bullet mass estimated by \citet{Bac90} using an assumption of $T_{\rm ex}{\sim}$20 K.
The momentum and kinetic energy of the bipolar jet are estimated to be 0.3 $M_{\odot}$ km s$^{-1}$ and 5.0$\times$10$^{44}$ erg, respectively.
Here, the calculation was done using the de-projected jet velocity under the assumption that the jet axis is inclined by 21$^{\circ}$ from the plane of the sky \citep{Gir01}.
The dynamical timescale of the jet derived from the length ($\sim$ 20\arcsec = 5000 AU) and mass weighted mean (de-projected) velocity is estimated to be only $\sim$150 yr.
Because of the rather large mass and short timescale, the obtained mass loss rate is also large, $\sim$10$^{-5}$ $M_{\odot}$ yr$^{-1}$.
Furthermore, the high velocity of $\sim$160 km s$^{-1}$ (corrected for the inclination of 21$^{\circ}$) brings extremely large momentum supply rate of $\sim$2${\times}$10$^{-3}$ $M_{\odot}$ km s$^{-1}$ yr$^{-1}$ and mechanical luminosity of $\sim$26$L_{\odot}$ for the jet.

It should be noted that the derived total mechanical luminosity of $\sim$26 $L_{\odot}$ is a factor of 3.5 larger than the bolometric luminosity of the central source (7.5 $L_{\odot}$).
One possible reason for such a discrepancy is that the mass of the jet was overestimated.
Since the CO $J$=3--2 emission comes from both the jet and shell, the high velocity part of the shell may contaminate the CO flux that was used to calculate the mass of the jet.
However, this effect should not be significant, because the CO flux from the shell does not dominate  the CO flux at ${\Delta}V >$50 km s$^{-1}$ velocity range.
The second possible reason is the excitation temperature.
The assumed excitation temperature of $T_{\rm ex}{\sim}$100 K is much lower then the kinetic temperature of $>$500 K derived from the far infrared measurements \citep{Nis99,Nis00}.
However, the assumption with higher excitation temperature increases the mass and dynamical parameters (i.e. momentum, kinetic energy, mass loss rate, momentum rate, and mechanical luminosity).
For example, the assumption of $T_{\rm ex}$= 500 K increases the mass and dynamical parameters by a factor of 3.8, and brings an extremely large mechanical luminosity of $\sim$90 $L_{\odot}$.
This implies that the bulk of the molecular gas in the jet is not so warm as 500 K, and that the warm gas which contributes to the far infrared emission is not the major component.
On the other hand, the lower excitation temperature of 40--50 K reduces the mass.
However this effect is limited to a factor of 1.5 at the most.
Another possible reason is the CO abundance, which was assumed to be 10$^{-4}$.
This value can be as high as 4$\times$10$^{-4}$ in the chemical model of protostellar winds proposed by \citet{Gla91}, in which molecules such as CO and SiO are formed via gas-phase reactions in an initially atomic protostellar wind.
If the higher CO abundance of 4$\times$10$^{-4}$is adopted, the mass and all dynamical parameters are reduced by a factor of 4, and the mechanical luminosity of the jet becomes $\sim$7 $L_{\odot}$, which is comparable to the bolometric luminosity of the central source.
However, this value for the mechanical luminosity of $\sim$7 $L_{\odot}$ should be the lower limit, because it is derived under the assumption of the optically thin CO emission.
The physical parameters derived using $T_{\rm ex}$ = 100 K and CO/H$_2$ abundance ratio of 4$\times$10$^{-4}$ are given Table 1.

\subsection{Physical parameters of the outflow shell}

The mass, momentum, and kinetic energy of the outflow shell were estimated using the CO flux measured in the velocity ranges of $\pm$1--40 km s$^{-1}$ from the systemic velocity.
The CO emission is assumed to be optically thin and in the LTE condition with an excitation temperature of 40 K, which is derived from the observed peak brightness temperature of the CO $J$=3--2 line.
Since most of the gas in the shell component is considered to be the swept-up ambient material, the canonical CO abundance of 10$^{-4}$ was used for the calculations.
The dynamical parameters of the outflow shell are summarized in Table 2. 
Since significant fraction of the CO flux is missed in the low velocity ranges, Table 2 show the parameters with and without correction for the effect of missing flux. 
The inclination angle of the outflow axis is assumed to be 21$^{\circ}$ as in the case of the EHV jet.
The dynamical timescale of the outflow shell is adopted to be $\sim$240 yr, which is derived from the modeling descried in the previous section.
Table 2 shows that the dynamical parameters of the outflow shell are comparable to those of the EHV jet listed in Table 1.
If the effect of the missing flux is corrected, the outflow shell has a momentum supply rate of  $\sim$1.8$\times$10$^{-3}$ $M_{\odot}$ km s$^{-1}$ yr$^{-1}$, and a mechanical luminosity of $\sim$8.6 $L_{\odot}$.
The mechanical lumiosity derived here is comparable to the bolometric luminosity of the central source and the mechanical luminosity of the EHV jet.
It should be noted that the dynamical parameters for the redshifted outflow are affected by the contamination from the L1448C(S) outflow (see the next section).
However, the effects of L1448C(S) outflow to the dynamical parameters are not so significant, because this outflow is seen only in the low velocity ranges.

\subsection{CO outflow from L1448C(S)}

In the CO maps in the low velocity ranges such as Fig. 4a and 4b, there is a  blueshifted component in the redshifted lobe.
This blueshifted component shows a triangle shape with its apex at the position of L1448C(S), and extends to the northeast direction with a position angle of $\sim$40$^{\circ}$.
It is likely that this blueshifted component is related to the activity of L1448C(S).
Fig.~\ref{fig11}, which provides a close-up view of the L1448C(S) region, shows a redshifted counterpart to the southwest of L1448C(S), a compact component at $\sim$1\arcsec southwest of L1448C(S) and another extended component to the southwest of the V-shaped shell of the L1448C(N) outflow.
This NE-SW outflow from L1448C(S) is also highly collimated.
The opening angle of the lobes is $\sim$40$^{\circ}$, which is similar to the redshifted lobe of the L1448C(N) outflow.
However, this outflow is seen only in the velocity ranges slower than 15 km s$^{-1}$, and has no high-velocity jet-like component.
There is no SiO $J$=8--7 emission either.

The redshifted part of the L1448C(S) outflow overlaps with the western wall of the L1448C(N) outflow.
At the place where the two outflows are superposed, the CO emission is significantly enhanced and the wall of the L1448C(N) outflow lobe is bending.
Therefore, it is possible that the two outflows are intersecting, although three dimensional geometries of two outflows are uncertain.

The physical parameters of this outflow were estimated assuming that the CO emission is optically thin and that the excitation condition of the CO is in LTE.
The fractional abundance of the CO was adopted to be the canonical value, 10$^{-4}$, because the V-shaped morphology and low velocity suggest that the bulk of the L1448C(S) outflow is likely to be the swept-up ambient gas.
A mean atomic weight of the gas was assumed to be 1.41.
The excitation temperature was assumed to be 20 K, which is same as the rotational temperature of the dense gas surrounding L1448C derived from the NH$_3$ observations \citep{Cur99}.
The mass of the blueshifted component is estimated to be 5.3$\times$10$^{-4}$ $M_{\odot}$ and that of the redshifted component is 2.7$\times$10$^{-4}$ $M_{\odot}$.
Since the CO $J$=3--2 emission is assumed to be optically thin, the values derived here are the lower limits.
In addition, the mass of the redshifted component is likely to be underestimated, because we exclude the region where the two outflows are superposed.
It is also possible that part of the flux from the spatially extended component is missing.
The dynamical parameters of the flow are summarized in Table~\ref{table3}.
Table 3 gives two kinds of values; one is corrected for the inclination effect  and the other is without this correction (uncorrected).
Here, the inclination angle of the L1448C(S) outflow from the plane of the sky was assumed to be  32.7$^{\circ}$, which corresponds to the mean inclination angle from assuming randomly oriented outflows.
We also assumed that the outflow lobes have a size of $\sim$20\arcsec (5000 AU).
The actual size of this outflow is not clear because of confusion with L1448C(N) outflow and its sidelobes.
However, there is no counterpart of this outflow in the single-dish CO $J$=3--2 map \citep{Hat07}.
This suggests that the spatial extent of this outflow is not large, or the spatially extended component is fainter than the sensitivity limit of the single-dish measurement. 
In the later case, the faint component would not make significant contribution to the dynamical parameters even though its spatial extension is large.

\section{DISCUSSION}

\subsection{Compact disk around L1448C(N)}

The compact component of dust continuum emission is partially resolved by the 0\farcs7$\times$0\farcs5 beam.
The observed structure is elongated perpendicular to the outflow axis, suggesting that this component traces the disk surrounding the protostar. 
If the beam-deconvolved major axis, $\sim$90 AU, represents the diameter of the disk, the disk size of L1448C(N) is comparable to those around the youngest protostars such as HH211 \citep{Lee07b} and HH212 \citep{Cod07, Lee07a, Lee08}.
If the measured flux (330 mJy) comes from the region with 0\farcs37$\times$0\farcs26 size, the brightness temperature corresponds to $\sim$35 K.
The spectral energy distribution (SED) of the compact source from 8.3 GHz to 350 GHz is shown in Fig.~\ref{fig12}.
The measured flux densities are fit by a single power law with a spectral index  $\alpha$ of 1.98 (solid line in Fig~\ref{fig12}), which agrees with the previous result ($\alpha$ = 1.84) of \citet{Sch04}.
This spectral index is smaller than $\alpha$ = 3.4 derived from the photometric broadband measurement including the contribution of the larger scale envelope component \citep{Fro05}.
Since the emission at cm wavelengths is likely to be the free-free emission from shock-ionized gas,
the contribution of the free-free component at mm and sub-mm wave ranges was estimated by using the equations given by  \citet{Cur90}.
Using the parameters of the stellar wind in \citet{Cur90}, the flux densities of the free-free emission at mm and sub-mm wave ranges were estimated to be less than 1 mJy (dotted line in Fig~\ref{fig12}).
Therefore, the small index number is unlikely to be due to the contribution of the free-free component.
The power law fit assumes that the emission is optically thin and that the Rayleigh-Jeans approximation is valid.
However, the observed index $\alpha\sim$2 is close to the value for the blackbody radiation, suggesting that the emission is optically thick.
In addition, a brightness temperature of 35 K suggests that the Rayleigh-Jeans approximation is not applicable in the mm and sub-mm wave ranges.
Therefore, we applied an optically thick fit without Rayleigh-Jeans approximation using the formula,
\begin{equation}
S_{\nu}={\Omega}_s B_{\nu}[1-{\rm exp}(-{\tau}_{\nu})],
\end{equation}
Where $S_{\nu}$ is the flux density, ${\Omega}_s$ is the source size, B$_{\nu}$ is the Planck function, and ${\tau}_{\nu}$ is the dust optical depth, which is assumed to follow the power low, ${\tau}_{\nu} {\propto} {\nu}^{\beta}$.
Assuming that the source size is 0\farcs37$\times$0\farcs26 and a dust temperature of 40 K, the SED fit provides $\beta$ = 1.3 and ${\tau}_{350 {\rm GHz}}$ = 7.5 (dash-dotted line in Fig~\ref{fig12}).
The average optical depth at 350 GHz for a disk of mass (gas+dust) $M_D$ and radius $R_D$ is given by 
\begin{equation}
<\tau_{350 {\rm GHz}}> = \left( \frac{0.5}{{\rm cos}  \theta} \right) \left( \frac{M_D}{0.1 M_{\odot}} \right) \left( \frac{R_D}{100 {\rm AU}} \right) ^{-2},
\end{equation}
where $\theta$ is the disk inclination angle to the line of sight \citep{Jor07}.
Using this relation, the mass of the disk with ${\tau}_{350 {\rm GHz}}$ = 7.5, $R_D$ = 45 AU, and $\theta$ = 69$^{\circ}$ (assuming that the disk is perpendicular to the jet axis) is estimated to be 0.11 $M_{\odot}$.
The mass derived here is approximately twice as large as the lower limit of 0.047 $M_{\odot}$, which is derived under the assumption of optically thin emission.

\subsection{Stellar mass loss rate and its implication to the protostellar evolution}

Since protostellar jet is considered to be closely linked to the mass accretion,  the stellar mass-loss rate gives us a rough estimate of the mass accretion rate onto the star.
Theoretical estimate for the ratio of mass outflow to mass accretion rate ($\dot{M}_{\rm out}/\dot{M}_{\rm acc}$) is ${\sim}$1/3 for an X-wind type magneto-centrifugal wind \citep[e.g.][]{Shu94}.
If the $\dot{M}_{\rm out}/\dot{M}_{\rm acc}$ ratio is assumed to be $\sim$0.3, the total mass-loss rate (blue + red) derived from the  CO flux, 2.4$\times$10$^{-6}$ $M_{\odot}$ yr$^{-1}$, gives us the mass accretion rate of  8$\times$ 10$^{-6}$ $M_{\odot}$ yr$^{-1}$.
In spite of rather high accretion rate, the observed bolometric luminosity is only 7.5 $L_{\odot}$, suggesting that the mass of the central star is still very small.
If most of the observed bolometric luminosity is released by means of accretion, the mass of the central star is calculated by using the relation, $M_*$ = $L_{\rm acc}R_{*}/G\dot{M}_{\rm acc}$, where $M_*$ is the mass of the central star, $L_{\rm acc}$ is the accretion luminosity, and $\dot{M}_{\rm acc}$ is the mass accretion rate onto the protostar.
The radius of the protostar, $R_*$, is considered to be $\sim$1 $R_{\odot}$ in the earliest evolutionary stage with very low mass and $\sim$3 $R_{\odot}$ in the later stage \citep[e.g.][]{Sta88}.
The mass of the central star is estimated to be 0.03--0.09 $M_{\odot}$ for an accretion luminosity of 7.5 $L_{\odot}$.
With a stellar mass of 0.03--0.09 $M_{\odot}$, the Keplerian velocity at the surface of the protostar becomes $\sim$80 km s$^{-1}$.
In this case, the jet velocity to Keplerian velocity ratio becomes $\sim$2, which is reasonable if the jet is launched by magneto-centrifugal force \citep{Shu94,Pud07}.
If a constant mass accretion rate of 8$\times$ 10$^{-6}$ $M_{\odot}$ yr$^{-1}$ is assumed, the age of the central star is estimated to be (4--12)$\times$10$^3$ yr.
The timescale derived here is consistent to the kinematic age of the larger scale outflow of $\sim$0.3 pc scale \citep{Bac90}.
However, morphology of the EHV jets implies that the mass accretion was variable.
The highly-colimated EHV jets terminate at $\sim$20\arcsec from the source, suggesting that L1448C(N) experienced lower activity phase in the past and enhanced its activity significantly in the last $\sim$150 yr.

\subsection{Clumpy structure in the L1448C(N) jet}

High resolution SiO and CO maps show that the EHV bullets B1 and R1 identified by \citet{Bac90} consist of chains of knots.
If the BI-II and RI-II knots are included, the knots are aligned with almost equal intervals of $\sim$2\arcsec (500 AU).
Similar knotty structure with semi-regular intervals is also seen in the SiO and CO jets in HH211 \citep{Hir06, Pal06, Lee07b} and HH212 \citep{Cod07, Lee07a}.
In the case of HH211 and HH212, the knots seen in the SiO and CO have their counterparts in the near infrared H$_2$ emission except the innermost knots pairs that were highly obscured.
In the case of the L1448C(N) outflow, H$_2$ emission knots are seen only in the northern blueshifted side and not in the southern redshifted side \citep{Dav94, Eis00}.
This is probably because the axis of the L1448C(N) jet is inclined from the plane of the sky and the near infrared emission in the southern part is obscured by the dense gas envelope traced by the NH$_3$ emission \citep{Cur90}.
In the northern side, the morphology of the SiO jet coincides well with that of the H$_2$ jet \citep{Eis00}, which also shows a kink between BI and BII components.
Therefore it is likely that the knots in the L1448 jet are the internal bow shocks in the jet beam as in the cases of HH211\citep{Hir06, Pal06, Lee07b} and HH212 \citep{Cod07, Lee07a}.
In fact, some of the SiO knots are partially resolved in the transverse direction, and the RII-a knot shows an arc-shaped structure typical of a bow shock (Fig. 3f).
The SiO emission is weak at the positions of the BI-II and RI-II knots, suggesting that the shocks at these positions are rather weak as compared to the other knot positions.
Since the jet is deflected at the positions of BI-II and RI-II knots, it is likely that the jet material there is impacting less dense material surrounding the jet beam.

\subsection{Jet bending}

As shown in Fig. 5,  the blue part and the red part of the jet are misaligned by  $\sim$5$^{\circ}$ and forming a C-shaped structure bending toward the west.
Such a C-shaped bending of the jet could be due to the Lorenz forces between the jet and interstellar magnetic field \citep{Fen98}, the orbital motion of the jet source in a binary system \citep{Fen98, Mas02}, or dynamical pressure from external medium \citep{Fen98}. 
In the case of the Lorenz forces, a C-shaped bending is expected if the poloidal current in the jet and counter jet flows in the same direction \citep{Fen98}.
However, this mechanism is difficult to account for the observed bending of the L1448C(N) jet, because typical interstellar magnetic field with several tens of microgauss is not strong enough to bend the jet beam with a density of $>$10$^6$ cm$^{-3}$.

If the C-shaped bending is produced by the orbital motion of a binary system, the orbital radius and orbital velocity can be estimated by using the analytical model of \citet{Mas02}.
Here the jet is assumed to be ejected at a velocity of $v_j$ from one of the binary protostars in a circular orbit of radius $r_0$ and orbital velocity $v_0$.
The $z$-axis is parallel to the orbital rotation axis, and $z$=0 is the orbital plane.
As shown in Fig. 3 of \citet{Mas02}, the deflection angle $\alpha$ of the jet beam near the source is approximated by the $x$=${\kappa}z/{\rm cos}i$ line, where $\kappa$=$v_0/v_j$ and $i$ is the inclination angle of the orbital axis with respect to the plane of the sky.
In the case of the L1448C(N) jet, the deflection angle $\alpha$ is estimated to be 2.5$^{\circ}$, which corresponds to the half of the misalignment angle.
Using the jet velocity of $\sim$160 km s$^{-1}$, the orbital velocity is calculated to be 6.5 km s$^{-1}$.
Since the mass of the protostar with jet is only 0.03--0.09 $M_{\odot}$,  the total mass of the binary system is considered to be less than 0.18 $M_{\odot}$.
Therefore, the radius and period of the orbital morion are estimated to be smaller than 4.2 AU and 20 yr, respectively.
However, such a short period orbital motion cannot account for the observed C-shaped pattern, because
the C-shaped bending is seen in the BI and RI parts of the jet with a length of $\sim$2000 AU with a dynamical time scale of  47 yr.
In order to produce the C-shaped pattern with the orbital motion, the orbital period needs to be longer than twice of the dynamical time scale.
 
In the case of dynamical pressure of external medium, ambient gas with $n$(H$_2$)$<$10$^4$ cm$^{-3}$ cannot account for bending the jet with a density of  $>$10$^6$ cm$^{-3}$,  unless the protostar is moving with a velocity that is comparable to the jet velocity.
On the other hand, the dynamical pressure caused by the outflow from the nearby protostar, L1448N, cannot be ruled out.
As shown in the CO map of \citet{Bac90}, the redshifted lobe of the L1448N outflow overlaps with the blue lobe of the L1448C(N) outflow.
The interaction between the two outflows from L1448C(N) and L1448N has been suggested by \citet{Bac95}, because the large scale outflow from L1448C(N) shows a considerable bending at the place where the two outflows are overlapping.
Since the redshifted emission from L1448N outflow reaches close to the position of L1448C(N) \citep{Bac95}, it is possible that the jet from L1448C(N) is propagating under the influence of the L1448N outflow.
In this case, the dynamical pressure from L1448N outflow acts from north to south, and deflects the jet beams to the west if they were ejected to the the northwest and southeast directions. 

\subsection{Deflection and wiggling of the jet}

In addition to the C-shaped bending, the jet is also deflected toward the east by $\sim$40$^{\circ}$ at the position of BI-II and toward the south by $\sim$15$^{\circ}$ at the RI-II position.
Since both sides of the jet are deflected at almost same distance from the central star, the jet deflection is likely to be caused by some variability intrinsic to the driving source rather than by external perturbation.
The observed morphology is similar to the S-shaped point-reflection symmetric pattern that is expected if the disk is  precessing or wobbling.
Although the jet is not exactly the S-shape but asymmetric in deflection angle, this is probably because of the projection effect.
In a binary protostellar system with a disk misaligned with the orbital plane of the binary, the disk wobbles with a period approximately half of the binary orbital period and precesses with a period of $\sim$20 orbital period \citep{Bat00}.
S-shaped point symmetry will be observed if the precession or wobbling time scale is longer than four times of the dynamical timescale.
Since the jet deflection occurs at BI-II and RI-II the time scale of which is $\sim$50 yr, the time scale of the precession or wobbling should be longer than $\sim$200 yr.
Therefore, if the deflection is due to the precession, the lower limit of the binary orbital period is $\sim$10 yr.
Since this orbital period of $\sim$10 yr is comparable to the period of small scale wiggling  shown in Fig. 6, 15--20 yr, this orbital motion can also explain the wiggling feature.
If the binary system consists of equal mass protostars with 0.03--0.09 $M_{\odot}$, the  orbital radius is estimated to be 2.4--4.2 AU.
On the other hand, if the jet deflection is due to the wobbling of the disk, the orbital period and the separation of the binary are estimated to be $\sim$400 yr and 30 AU, respectively.
Since the estimated separation of the binary is smaller than the size of the disk observed with the 350 GHz continuum emission,  it is possible that the observed 90 AU scale disk harbors two sources separated by 60 AU.
However, the binary with a separation of $>$60 AU cannot account for the small scale wiggling feature.

\subsection{Velocity variation of the jet}

The P-V diagram of the SiO shows that the velocity of the jet varies semi-periodically.
The velocity variation is more obvious in Fig.~\ref{fig13}, which plots the velocity centroid of the SiO emission in the redshifted part of the jet as a function of the distance from L1448C(N).
It is shown that the typical amplitude of the variation in velocity centroid is $\sim$7 km s$^{-1}$.
Such a velocity variation is expected if the jet is precessing, the jet is launched from an orbiting object,  or the ejection velocity itself varies as a function of time \citep[e.g.][]{Smi97}.
The period of the velocity variation estimated from the de-projected jet velocity and knot separation is $\sim$15--20 yr.
Since this time scale is much shorter than the precession time scale that is estimated to be 200 yr, it is unlikely that the velocity variation is caused by the precession of the jet.
On the other hand, the orbital motion of the driving source can account for the velocity variation; the binary system with an orbital period of $\sim$15--20 yr, an orbital radius of 2.4--4.2 AU, a total mass of 0.06--0.18 $M_{\odot}$ has an orbital velocity of $\sim$4.7--6.2 km s$^{-1}$, which is comparable to the amplitude of the radial velocity variation.
However, the orbital motion cannot explain the relation between the SiO intensity and velocity gradient.
As shown in the P-V map (Fig.~\ref{fig7}), each SiO knot has its higher velocity in the upstream side and lower velocity in the downstream side.
The opposite velocity gradient is always seen in the faint emission between the knots.
Such a structure is more likely to be formed by means of periodic variation of the ejection velocity. 
In such a case, the SiO knots are considered to be formed as the fast moving material plunge into the slow moving material in the downstream \citep[e.g.][]{Sto93, Sut97}.
The periodic variation in the ejection velocity is probably due to the modulation of mass accretion by means of companion.
In such a case, the variation amplitude of the jet velocity corrected for the inclination is calculated to be $\sim$20 km s$^{-1}$. 
This velocity amplitude corresponds to the shock velocity, which is consistent to the velocity of C-type shocks that can account for the excitation conditions of the far infrared molecular lines \citep{Nis99, Nis00}.
Similar velocity gradients in the knots with the faster part in the upstream side and the slower part in the downstream was also observed in the optical jet of HH111 \citep{Rag02} and in the CO and SiO jets from IRAS 04166+2706 \citep{San09}.

\subsection{Driving mechanism of the CO outflow}

The P-V diagram of the CO $J$=3--2 along the axis (Fig.~\ref{fig7}) exhibits two kinematic components, i.e. the EHV jet with an almost constant velocity and the outflow shell with a parabolic velocity pattern.
Although highly-collimated jet is clearly seen in both SiO and CO, the parabolic velocity pattern seen in the outflow shells is reproduced by the wind-driven model.
Therefore, the observational results require a wide-opening angle wind and a collimated jet at the same time.

One possible mechanism to explain the observed jet+shell structure is the ^^ ^^ unified model'' proposed by \citet{Sha06}, in which highly-collimated jet component is explained as an on-axis density enhancement of the {\it X}-wind type of wide opening angle wind launched magnetocentrifugally.
In this model, the jet along the axis corresponds to the densest part of the primary wind, and the shell is mostly consisted of the swept-up ambient material.
Their numerical model successfully reproduced the structure of a dense and narrow jet surrounded by a conical shell. 
The other models that can explain the two components structure are proposed by \citet{Mac08} and \citet{Ban06}.
The model proposed by \citet{Mac08} predicts two distinct flows from the adiabatic core and the protostar.
The flow from the adiabatic core driven by the magnetocentrifugal mechanism has a low velocity and a wide opening angle, while the flow from the protostar, which is mainly driven by the magnetic pressure gradient force, has a high velocity and is well collimated.
On the other hand, a model proposed by \citet{Ban06} predicts the structure with the jet powered by magnetocentrifugal force enclosed by the large-scale outflow driven by the magnetic pressure.
Although these two models reproduce the jet+shell structure similar to the observational results, the velocities of the shells predicted in these model are rather small ($\sim$5 km s$^{-1}$ for the model of \citet{Mac08}) because of the shallow gravitational potential at the launching point.
In the case of the L1448C(N) outflow, the terminal velocity of the outflow shellreaches to ${\Delta}V {\sim} {\pm}$70 km s$^{-1}$ without inclination correction, which is comparable to the velocity of the EHV jet.
In order to launch such a high velocity wind, the launching point of this wind should be close to the launching point of the EHV jet.
Therefore, the models with two components launched from two different regions do not explain the jet+shell structure in the L1448C(N) outflow.
In the case of the {\it X}-wind with density stratification \citep{Sha06}, the rather high velocity in the shell component is naturally explained because the shell is driven by the high-velocity primary wind launched at the same region as the EHV jet.

\subsection{Origin of the SiO in the jet}

It is considered that the SiO molecules observed in jets and outflows are formed as a consequence of grain sputtering in a C-shock releasing Si-bearing material into the gas phase, followed by the reaction with O and OH \citep{Sch97, Cas97, Gus08a}.
The multi-transition SiO lines from the L1448C(N) jet observed by \citet{Nis07} have been successfully modeled by the steady-state C-shock model of  \citet{Gus08a}  with a pre-shock density of $\sim$10$^5$ cm$^{-3}$ and a shock velocity of 30--45 km s$^{-1}$.
However, the conversion of Si into SiO is initially rather slow in their models, and the predicted  SiO line emission predominantly arises from postshock gas  $>$100 yr after the passage of shocks.
Since most of the SiO knots in the L1448C(N) jet have dynamical timescale shorter than the SiO formation timescale, the steady-state C-type shock models of \citet{Gus08a} does not account for the SiO in the knots close to the central star, especially in the innermost knot pair with extremely short time scale less than 10 yr.
One possible explanation is that the SiO molecules existed on the grain mantles and are released into the gas-phase by means of shocks as suggested by \citet{Gus08b}.

Another possibility is the formation of SiO in high density primary jet \citep{Sha06}.
\citet{Gla89, Gla91} studied the formation of molecules in protostellar winds, which are originally neutral atomic, and found that significant quantities of SiO can be formed quickly in the close vicinity ($<$0.1 AU) of the central star if the mass-loss rate is high ($>$10$^{-6}$ $M_{\odot}$).
Since the mass-loss rate of the L1448C(N) jet is high enough, this scenario of in situ formation also can be the origin of the SiO in the jet.
The chemical models of \citet{Gla89, Gla91} also predict significant amount of CO synthesized in the winds; the CO abundance reaches an equilibrium value of 4$\times$10$^{-4}$ under the conditions in which observable amount of SiO is formed.
The morphological and kinematical similarity of the CO and SiO jets supports the idea that both CO and SiO are formed in the protostellar wind.

\subsection{Properties of L1448C(S)}

The secondary source L1448C(S) is located at $\sim$8\farcs3 (2000 AU) southeast of L1448C(N).
The 350 GHz continuum flux of this source is $\sim$60 mJy, which is five times smaller than the flux from L1448C(N).
If an optically thin condition is assumed, the mass of the circumstellar material surrounding L1448C(S) is estimated from the observed flux using the dust mass opacity of 1.75 cm$^2$ g$^{-1}$ \citep{Oss94} and the formula given by \citet{Jor07}.
For a dust temperature of $\sim$40 K, the estimated mass in the optically thin limit is 8.6$\times$10$^{-3}$ $M_{\odot}$.
Since L1448C(S) is associated with a molecular outflow, it is highly probable that this source is a protostar rather than a mere dust clump at the cavity wall as claimed by \citet{Mau10}.
The NH$_3$ data of \citet{Cur99} suggest that L1448C(S) is formed in the same dense core as L1448C(N).
However, the SED of L1448C(S) is significantly different from that of L1448C(N).
In table 4, the broad-band spectra of L1448C(N) and L1448C(S) measured at different wavelengths are listed.
In the near-infrared, L1448C(S) is much dimmer than L1448C(N); only L1448C(N) appeared in the $K_s$ band image of \citet{Tob07}.
On the contrary, in the mid-inrfrared at three IRAC bands, band 2 (4.5 $\mu$m), band 3 (5.8 $\mu$m), and band 4 (8.0 $\mu$m), L1448C(S) becomes brighter than L1448C(N); especially in the bands 3 and 4, L1448C(S) is more than six times brighter than L1448C(N).
In the MIPS 24 $\mu$m image, L1448C(S) is also seen clearly  \citep{Tob07, Reb07}.
The flux from L1448C(S) at 24 $\mu$m looks similar to that from L1448C(N), although the accurate flux value of each source is not easy to measure because of the confusion.
In the sub-mm and mm wavebands, L1448C(S) is much weaker than L1448C(N).
The masses of the circumstellar material surrounding L1448C(S), $\sim$0.01 $M_{\odot}$, is approximately 10 times smaller than that of L1448C(N), $\sim$0.1 $M_{\odot}$.
Due to the small amount of circumstellar material, the central star of L1448C(S) is likely to be less obscured in the mid-infrared as compared to that of L1448C(N) enshrouded by the thick cocoon.
The outflow activities in two sources are also different significantly.
The CO outflow from L1448C(S) is compact and substantially weaker than the L1448C(N) outflow.
The momentum flux of the L1448C(S) outflow is only $\sim$10$^{-6}$ $M_{\odot}$ km s$^{-1}$ yr$^{-1}$, which is two or three orders of magnitude smaller than that of the L1448C(N) outfow and is comparable to those of class I outflows studied by \citet{Bon96}.
In addition, there is no EHV component nor SiO emission associated with the outflow from L1448C(S).

The small amount of circumstellar material of less than 0.01 $M_{\odot}$ suggests that L1448C(S) may have accumulated most of its circumstellar mass.
The less energetic outflow also support the idea that L1448C(S) has a nature close to class I rather than class 0.
These results imply that two sources, L1448C(N) and L1448C(S) are formed in different epochs in the same dense core.
Another possibility is the effect of the L1448C(N) outflow.
Since L1448C(S) is located at the same line of sight as the L1448C(N) outflow, it is possible that the outflowing gas has stripped away the dense gas surrounding L1448C(S).
Although the three-dimensional geometries of the sources and outflows are not clear, the high rotational temperature of NH$_3$ observed at the position of L1448C(S) \citep{Cur99} suggests the possibility that the gas around L1448C(S) is heated by the interaction with the jet from L1448C(N).
In such a case, the {\it apparent} age of L1448C(S) is older than that of L1448C(N) simply because the amount of material left around L1448C(S) is smaller than that around L1448C(N).
The effect of outflow is also proposed to explain the difference of the apaprent evolutionary stage of protostellar pair L1448N(A) and L1448N(B), which is located at $\sim$75\arcsec northwest of L1448C \citep{Oli06}.

The third scenario is the disintegration of an unstable multiple system as proposed by \citet{Rei00}.
Since nonhierarchical triple systems are unstable, they break up, ejecting the lightest member, while the remaining two members form a close binary system with a highly eccentric orbit.
In this scenario, the disks around escaping stars will be highly truncated; the typical disk size is expected to be around half of the distance between the stars in the close triple encounter.
If L1448C(S) is the escaping member, the small amount of its circumstellar material can be explained by means of disk truncation.
This scenario also implies that L1448C(N) is a close binary system.
The observed deflection, wiggling, and periodic velocity variation of the jet suggest the possibility that L1448C(N) is a close binary system with an orbital radius of $\sim$2--4 AU.
Therefore, it is possible that such a close binary system was formed by means of disintegration of a triple system.
In order to assess this scenario, kinematic information of L1448C(S) becomes important.
Although previous NH$_3$ results of \citet{Cur99} did not show peculiar motions in the dense core containing L1448C(N) and L1448C(S), a detailed study with higher angular resolution would be helpful.

\section{CONCLUSIONS}
The central region of the highly-colimated molecular outflow driven by L1448C was mapped in the SiO $J$=8--7, CO $J$=3--2, and 350 GHz continuum emission with the SMA at $\sim$1 arcsecond resolution.
Our main conclusions are the following:
\begin{enumerate}
\item The 350 GHz continuum emission was detected from two {\it Spitzer} sources L1448C(N) and L1448C(S). 
The continuum emission from L1448C(N) consists of an extended component  and a compact component. 
The compact component is elongated perpendicular to the outflow axis, and is likely to be a circumstellar disk with a size of $\sim$90 AU. 
The spectral index of this compact component derived from the data from 86 GHz to 350 GHz is ${\alpha}{\sim}$2, suggesting the possibility that the continuum emission is optically thick at 350 GHz.
The mass of the disk is estimated to be $\sim$0.11 $M_{\odot}$.
\item The continuum flux from L1448C(S) is $\sim$60 mJy, which is $\sim$10 times lower than the flux from L1448C(N), although L1448C(S) is brighter than L1448C(N) in the mid infrared wavebands.
The mass of the circumstellar material surrounding L1448C(S) is estimated to be 8.6$\times$10$^{-3}$ $M_{\odot}$.
\item A narrow jet from L1448C(N) along the outflow axis was observed in the SiO and the high-velocity CO. The width of the jet measured in the SiO images is $\sim$200 AU FWHM on average. 
The jet consists of a chain of emission knots with an inter-knot spacing of $\sim$500 AU.
It is likely that the knots in the L1448 jet are the internal bow shocks in the jet beam.
\item The dynamical timescale of the innermost pair of knots, which are significant in the SiO but barely seen in the CO, is only $\sim$10 yr. 
It is likely that the SiO may have been formed quickly in the protostellar wind through the gas-phase reaction, or been formed on the dust grain and directly released into the gas phase by means of shocks.
\item The low velocity CO emission delineates two V-shaped shells with a common apex at  L1448C(N). The kinematics of this shell component is reproduced by the model of wide opening angle wind. 
Therefore, the outflow from L1448C(N) consists of both highly-collimated jets and shells produced by wide-opening angle wind.
The observed jet+shell structure can be explained by the ^^ ^^ unified model'' proposed by \citet{Sha06}, in which highly-collimated jet components are explained as an on-axis density enhancement of the {\it X}-wind type of wide opening angle wind.
\item The Jet from L1448C(N) is extremely active with a momentum supply rate of $\sim$5$\times$10$^{-4}$ $M_{\odot}$ km s$^{-1}$ yr$^{-1}$ and a mechanical luminosity of $\sim$7 $L_{\odot}$. 
The mass accretion rate derived from the mass loss rate  is $\sim$10$^{-5}$ $M_{\odot}$ yr$^{-1}$.
Such a high mass-accretion rate and a rather low bolometric luminosity of the central source, 7.5 $L_{\odot}$, imply that the central protostar is still in the very early phase of its evolution with a mass of 0.03--0.09 $M_{\odot}$ and a dynamical age of (4--12)$\times$10$^3$ yr.
\item The blue part and the red part of the jet are misaligned by $\sim$5$^{\circ}$ and forming a C-shaped bending toward the west.
The possible origin of this bending is the dynamical pressure caused by the outflow from the nearby protostar, L1448N.
\item The jet is deflected toward the east in the blueshifted part and toward the south in the redshifted part. 
In addition, the jet is wiggling with a period of $\sim$15--20 yr.
The deflection and wiggling of the jet can be explained if the driving source is a member of the binary system with an orbital radius of 2--4 AU.
\item The jet shows a semi-periodic variation in radial velocity with an amplitude of $\sim$7 km s$^{-1}$.
Each SiO knot has its higher velocity in the upstream side and lower velocity in the downstream side. The opposite velocity gradient is seen in the faint emission between the knots.
It is likely that the ejection velocity varies periodically by means of modulation of mass accretion.   
\item The bipolar outflow in the NE-SW direction centered at L1448C(S) was discovered in the CO $J$=3--2. 
This provides a strong evidence that L1448C(S) is a protostar.
The momentum flux of this outflow is only $\sim$10$^{-6}$ $M_{\odot}$ km s$^{-1}$ yr$^{-1}$, which is two to three orders of magnitude smaller than that of L1448C(N) outflow, and is comparable to those of class I outflow.
\item L1448C(S) is surrounded by a rather small amount of circumstellar material of less than 0.01 $M_{\odot}$ and is powering a less energetic outflow, suggesting that this source has a nature close to class I rather than class 0, even though this source is formed in the same dense core as L1448C(N).
One possible scenario to explain this dichotomy is the effect of the outflow from L1448C(N); significant amount of material in the envelope surrounding L1448C(S) might have been stripped off by the powerful outflow from L1448C(N).
Another possibility is the disintegration of an unstable multiple system, in which L1448C(S) is an escaping member with a truncated disk.

\end{enumerate}

\acknowledgments

We wish to thank all the SMA staff in Hawaii, Cambridge, and Taipei for their 
enthusiastic help during these observations. 
N. Hirano thanks M. Machida and S. Inutsuka for fruitful discussion.
N. Hirano is supported by NSC grant 96-2112-M-001-023.

\clearpage

\begin{deluxetable}{lcc}
\tablecolumns{3}
\tablewidth{0pc}
\tablecaption{Dynamical parameters of the EHV jet from L1448C(N)}
\tablehead{
\colhead{Parameters} & \colhead{Blue}   & \colhead{Red}}
\startdata
Mass ($M_{\odot}$) & 2.3$\times$10$^{-4}$ & 2.2$\times$10$^{-4}$ \\
Momentum ($M_\odot$ km s$^{-1}$) & 0.038 & 0.037\\
Kinetic energy (erg) & 6.2$\times$10$^{43}$ & 6.2$\times$10$^{43}$ \\
Mean velocity (km s$^{-1}$) & 160.4 & 164.6 \\
Dynamical timescale (yr) & 160 & 150 \\
Mass loss rate\tablenotemark{a} ($M_\odot$ yr$^{-1}$) & 1.2$\times$10$^{-6}$ & 1.2$\times$10$^{-6}$ \\
Momentum supply rate ($M_\odot$ km s$^{-1}$ yr$^{-1}$) & 2.3$\times$10$^{-4}$ & 2.5 $\times$10$^{-4}$ \\
Mechanical luminosity ($L_\odot$) & 3.2 & 3.4 \\
\enddata

\label{table1}
\tablenotetext{a}{The primary jet velocity is assumed to be 200 km s$^{-1}$. }
\end{deluxetable}    

\clearpage                         

\begin{deluxetable}{lccccc}
\tablecolumns{6}
\tablewidth{0pc}
\tablecaption{Dynamical parameters of the L1448C(N) outflow shell}
\tablehead{
\colhead{}    &  \multicolumn{2}{c}{Blue} &   \colhead{}   &
\multicolumn{2}{c}{Red} \\
\cline{2-3} \cline{5-6} \\
\colhead{Parameters} & \colhead{uncorrected\tablenotemark{a}}   & \colhead{corrected\tablenotemark{b}}    &
\colhead{}    & \colhead{uncorrected\tablenotemark{a}}   & \colhead{corrected\tablenotemark{b}}}
\startdata
Mass ($M_{\odot}$) & 1.9$\times$10$^{-3}$ & 5.5$\times$10$^{-3}$ & & 3.0$\times$10$^{-3}$ & 8.6$\times$10$^{-3}$ \\
Momentum ($M_\odot$ km s$^{-1}$) & 0.10 & 0.18 && 0.15 & 0.26\\
Kinetic energy (erg) & 8.1$\times$10$^{43}$ &1.1$\times$10$^{44}$ && 1.0$\times$10$^{44}$ & 1.4$\times$10$^{44}$ \\
Momentum supply rate\tablenotemark{c} ($M_\odot$ km s$^{-1}$ yr$^{-1}$) & 4.1$\times$10$^{-4}$ & 7.4$\times$10$^{-4}$ && 6.1$\times$10$^{-4}$ & 10.9 $\times$10$^{-4}$ \\
Mechanical luminosity\tablenotemark{c} ($L_\odot$) & 2.8 & 3.8 && 3.5 & 4.8 \\
\enddata

\label{table2}
\tablenotetext{a}{The effect of missing flux is not corrected.}
\tablenotetext{b}{The effect of missing flux is corrected.}
\tablenotetext{c}{The dynamical time scale is assumed to be 240 yr.}
\end{deluxetable}

\clearpage

\begin{deluxetable}{lccccc}
\tablecolumns{6}
\tablewidth{0pc}
\tablecaption{Dynamical parameters of the L1448C(S) outflow}
\tablehead{
\colhead{}    &  \multicolumn{2}{c}{Blue lobe} &   \colhead{}   &
\multicolumn{2}{c}{Red lobe} \\
\cline{2-3} \cline{5-6} \\
\colhead{Parameters} & \colhead{{\it i}=32.7$^\circ$}   & \colhead{uncorrected}    &
\colhead{}    & \colhead{{\it i}=32.7$^\circ$}   & \colhead{uncorrected}}
\startdata
Mass ($M_\odot$) & \multicolumn{2}{c}{5.3$\times$10$^{-4}$} && \multicolumn{2}{c}{2.7$\times$10$^{-4}$} \\
Momentum ($M_\odot$ km s$^{-1}$) & 5.9$\times$10$^{-3}$ & 3.2$\times$10$^{-3}$ & 
& 2.0 $\times$10$^{-3}$ & 1.1$\times$10$^{-3}$ \\
Kinetic energy (erg) & 8.7$\times$10$^{41}$ &2.6$\times$10$^{41}$ & 
& 1.7$\times$10$^{41}$ & 5.0$\times$10$^{40}$ \\
Dynamical timescale (yr) & 2.5$\times$10$^{3}$ & 3.9$\times$10$^{3}$ & 
& 3.9$\times$10$^{3}$ & 6.1$\times$10$^{3}$ \\
Mass loss rate\tablenotemark{b} ($M_\odot$ yr$^{-1}$) & 2.4$\times$10$^{-8}$ & 8.2$\times$10$^{-9}$ & 
& 5.0$\times$10$^{-9}$ & 1.7$\times$10$^{-9}$ \\
Momentum supply rate ($M_\odot$ km s$^{-1}$ yr$^{-1}$) & 2.4$\times$10$^{-6}$ & 8.2$\times$10$^{-7}$ & 
& 5.0$\times$10$^{-7}$ & 1.7$\times$10$^{-7}$ \\
Mechanical luminosity ($L_\odot$) & 2.9$\times$10$^{-3}$ & 5.5$\times$10$^{-4}$ & 
& 3.6$\times$10$^{-4}$ & 6.7$\times$10$^{-5}$ \\
\enddata

\label{table3}
\tablenotetext{b}{The primary wind speed is assumed to be 100 km s$^{-1}$. Then we derived the mass loss rate by assuming momentum conservation of the the entrained CO gas and the primary wind.}
\end{deluxetable}    

\clearpage                                                                                       

\begin{deluxetable}{lccc}
\tablecolumns{4}
\tablewidth{0pc}
\tablecaption{Broad-band spectra of L1448C(N) and L1448C(S)}
\tablehead{
\colhead{}    &  \multicolumn{2}{c}{Flux (mJy)} &   \colhead{} \\
\cline{2-3} \\
\colhead{Wavelength} & \colhead{L1448C(N)}   & \colhead{L1448C(S)}    &
\colhead{Ref.} }
\startdata
1.65 $\mu$m & 0.144$\pm$0.016 & --- & 1 \\
2.1 $\mu$m & 1.07$\pm$0.13 & --- & 1 \\
3.6 $\mu$m & 4.2$\pm$0.4 & 2.5$\pm$0.08 & 2 \\
4.5 $\mu$m & 13.5$\pm$0.8 & 23.6$\pm$0.6 & 2 \\
5.8 $\mu$m & 11.7$\pm$0.8 & 77.2$\pm$0.8 & 2 \\
8.0 $\mu$m & 19.6$\pm$1.2 & 123$\pm$1 & 2 \\
860 $\mu$m & 330$\pm$6 & 60$\pm$6 & 3 \\
870 $\mu$m & 370$\pm$8 & 44$\pm$8 & 4 \\
1300 $\mu$m & 120$\pm$3 & 13$\pm$3 & 4 \\
\enddata

\label{table4}
\tablerefs{
(1) Tobin et al. 2007; (2) J{\o}rgensen et al. 2006; (3) This work; (4) J{\o}rgensen et al. 2007}
\end{deluxetable}    

\clearpage             

\begin{figure}
\epsscale{0.75}
\plotone{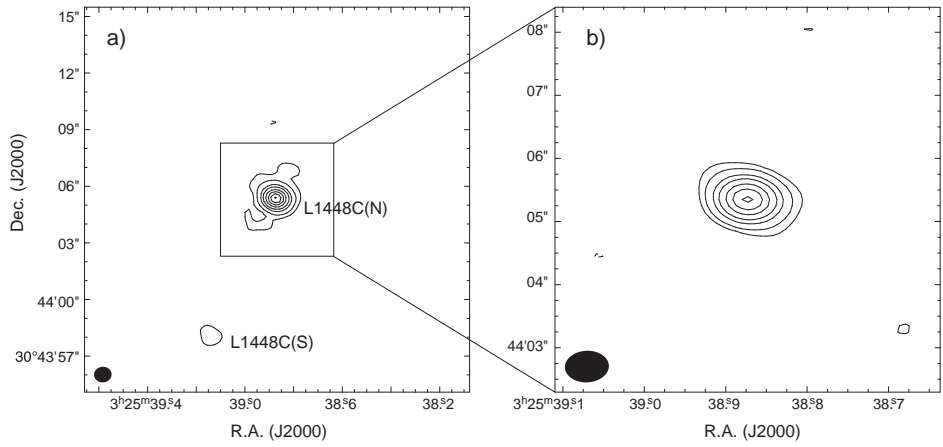}
\caption{350 GHz continuum emission from the central part of L1448-C. 
a) Map made using all the available visibility data. 
 The contours start at 3$\sigma$ and are drawn at 6, 12, 18, 27, 36, 45, and 54$\sigma$. 
 The 1$\sigma$ level is 6.42 mJy beam$^{-1}$.
b) Map made using the visibility data with the {\it uv} distance greater than 70 k$\lambda$. 
 The contours start at 3$\sigma$ and are drawn at 6, 12, 18, 27, 36, and 45$\sigma$. 
 The 1$\sigma$ level is 6.45 mJy beam$^{-1}$.
\label{fig1}}
\end{figure}

\clearpage 

\begin{figure}
\plotone{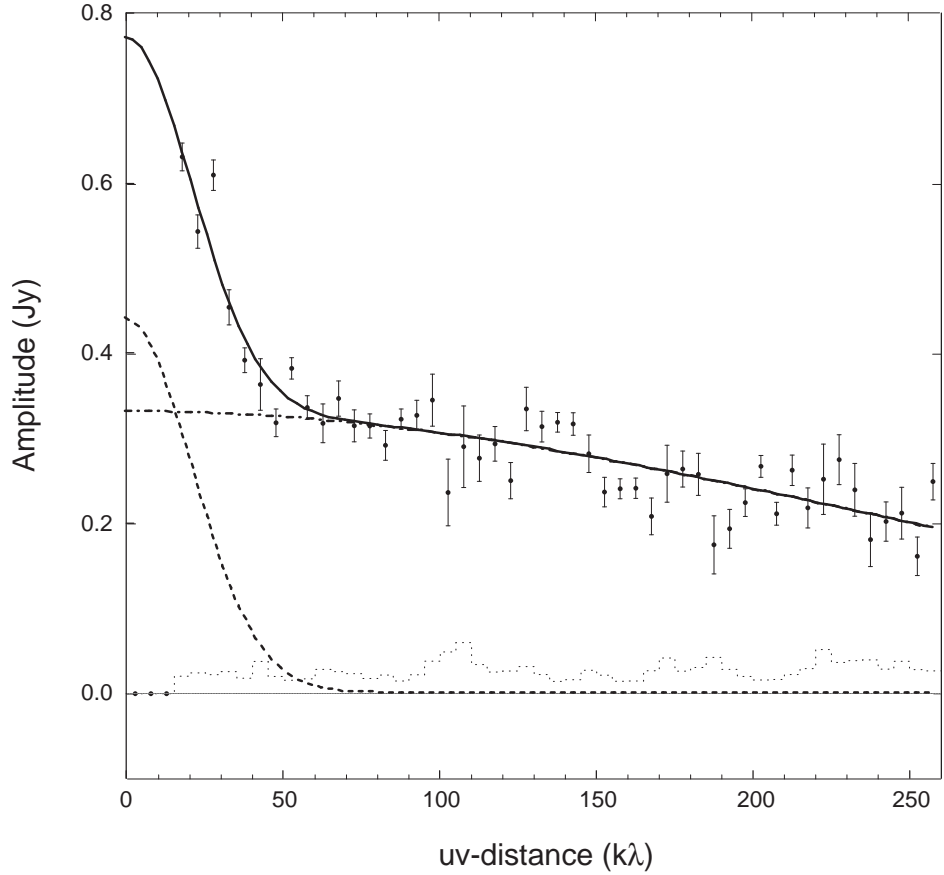}
\caption{Visibility amplitude versus {\it uv} distance plot for the continuum emission with 1 $\sigma$ error bars. The dotted histogram is the expected amplitude for zero signal. The profile can be fitted with two gaussian components. The dashed line is the curve for the extended component and the dash-dotted line is for the compact component. The solid curve is the total amplitude of the two components.
\label{fig2}}
\end{figure}

\clearpage 

\begin{figure}
%\plotone{fig3.eps}
\includegraphics[scale=.70]{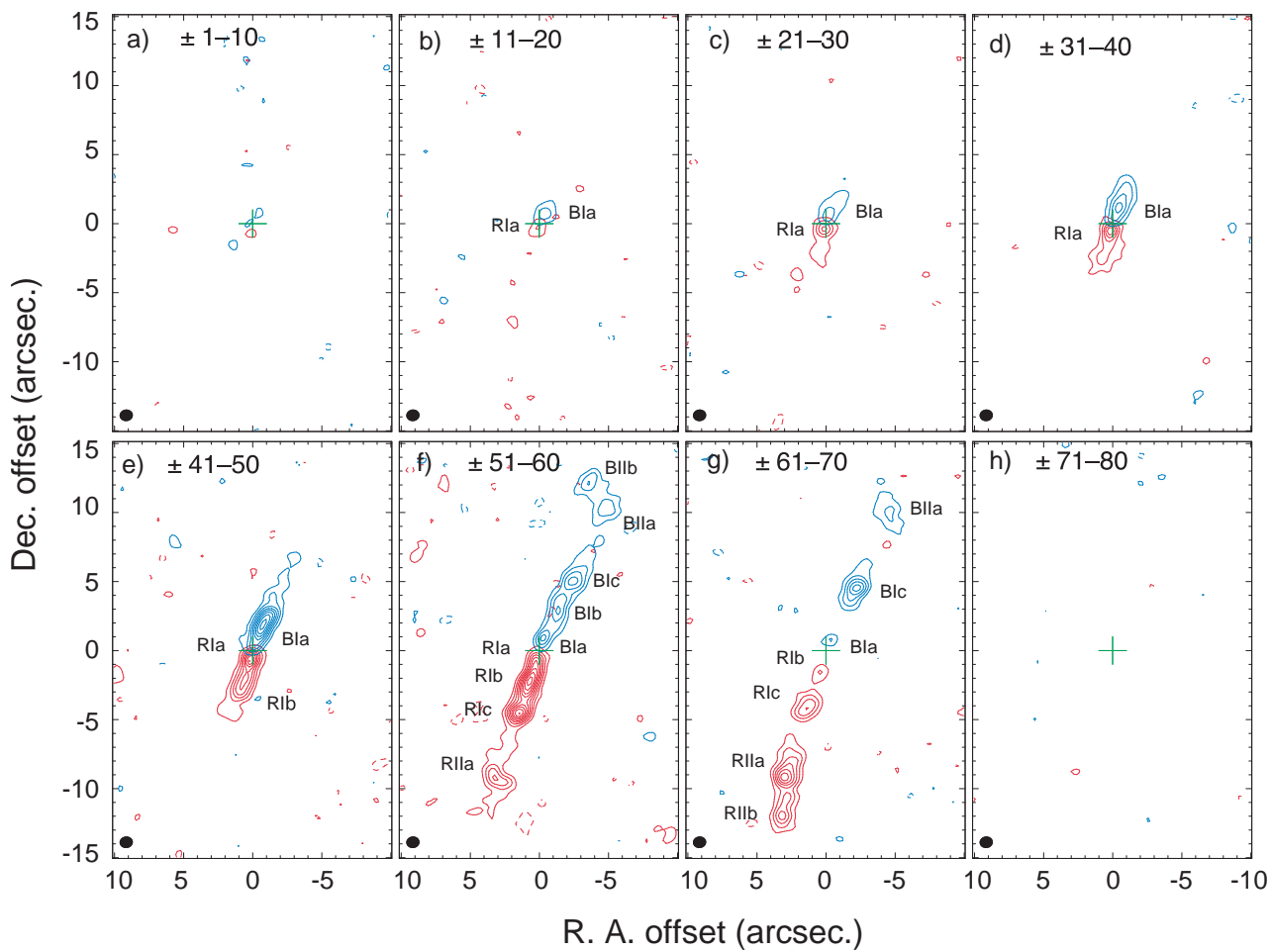}
\caption{Maps of the SiO $J$=8--7 emission averaged over 10 km s$^{-1}$ wide intervals. The velocity ranges with respect to the systemic velocity of $V_{\rm LSR}$ = 5.0 km s$^{-1}$ are shown in the upper sides of the panels. The cross marks the position of L1448C(N). The contour interval is 1.5 Jy beam$^{-1}$ km s$^{-1}$ (3 $\sigma$) with the lowest contour at 1.5 Jy beam$^{-1}$ km s$^{-1}$ (3 $\sigma$).
\label{fig3}}
\end{figure}

\clearpage 

\begin{figure}
%\plotone{fig4.eps}
\includegraphics[scale=.70]{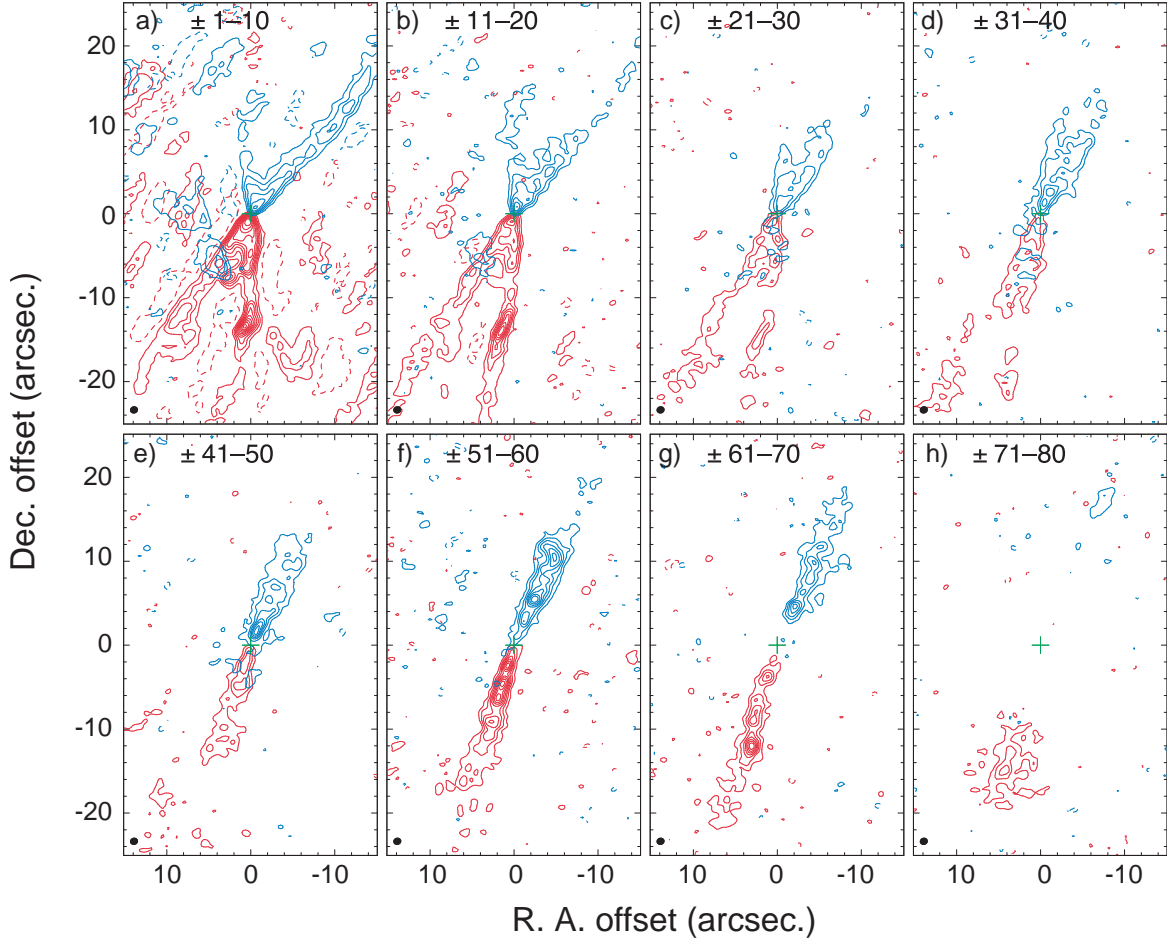}
\caption{Maps of the CO $J$=3--2 emission averaged over 10 km s$^{-1}$ wide intervals.   The velocity ranges with respect to the systemic velocity of $V_{\rm LSR}$ = 5.0 km s$^{-1}$ are shown in the upper sides of the panels. The cross marks the position of L1448C(N). 
a) The contours are drawn every 2.0  Jy beam$^{-1}$ km s$^{-1}$ (4 $\sigma$) with the lowest contours at 2.0 Jy beam$^{-1}$ km s$^{-1}$ (4 $\sigma$).
b)--h)  The contours are drawn every 1.5  Jy beam$^{-1}$ km s$^{-1}$ (3 $\sigma$) with the lowest contours at 1.5 Jy beam$^{-1}$ km s$^{-1}$ (3 $\sigma$).
\label{fig4}}
\end{figure}

\clearpage

\begin{figure}
\plotone{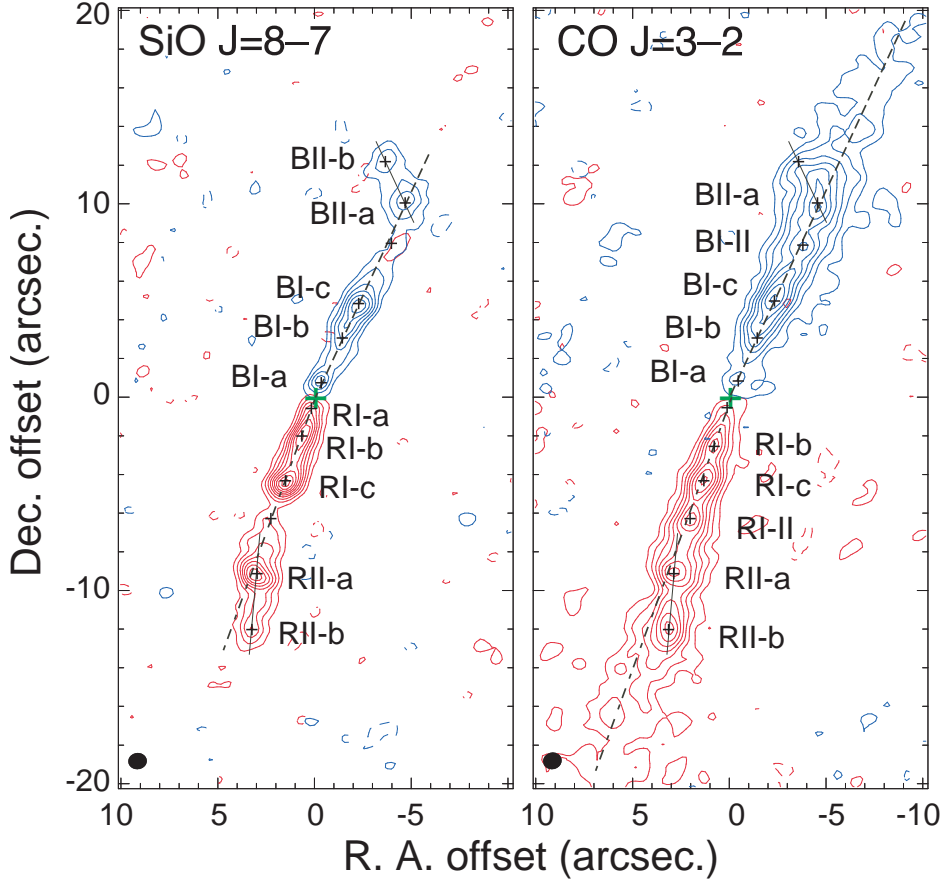}
\caption{High-velocity SiO ({\it left}) and CO ({\it right}) emission. The velocity ranges are $\pm$51--70 km s$^{-1}$ with respect to the systemic velocity. The contours are drawn every 2.01 Jy beam$^{-1}$ km s$^{-1}$ (3$\sigma$) with the lowest contour level at 2.01 Jy beam$^{-1}$ km s$^{-1}$ (3 $\sigma$). Green cross in each panel denotes the position of the continuum peak for L1448C(N). Black crosses mark the positions of the knots. Dashed straight line and dash-dotted line indicate the axes of BI (P.A. = $-$25$^{\circ}$) and RI (P.A. = $-$20$^{\circ}$) component, respectively. Thin solid lines are the axes of BII (P.A. = +15$^{\circ}$) and RII (P.A. = $-$5$^{\circ}$) components. }
\label{fig5}
\end{figure}

\clearpage

\begin{figure}
\plotone{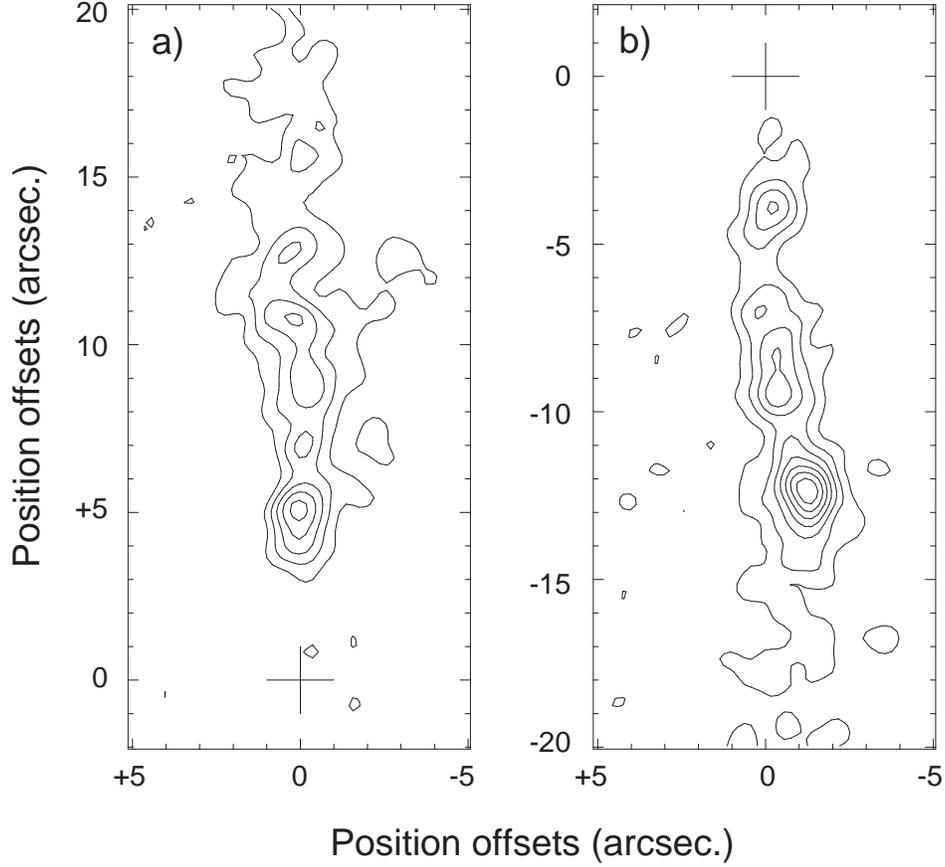}
\caption{(a) The CO $J$=3--2 jet in the velocity range from $-$70 to $-$61 km s$^{-1}$ with respect to the cloud systemic velocity. The cross marks the position of L1448C(N). The map is rotated by 25$^{\circ}$ counterclockwise. The contours are drawn every 1.5 Jy beam$^{-1}$ (3 ${\sigma}$) with the lowest contour of 1.5 Jy beam$^{-1}$ (3 ${\sigma}$) . (b) The CO $J$=3--2 jet in the velocity range from $+$61 to $+$70 km s$^{-1}$. The map is rotated by 20$^{\circ}$ counterclockwise. The contour levels are the same as (a).
}
\label{fig6}
\end{figure}

\begin{figure}
\plotone{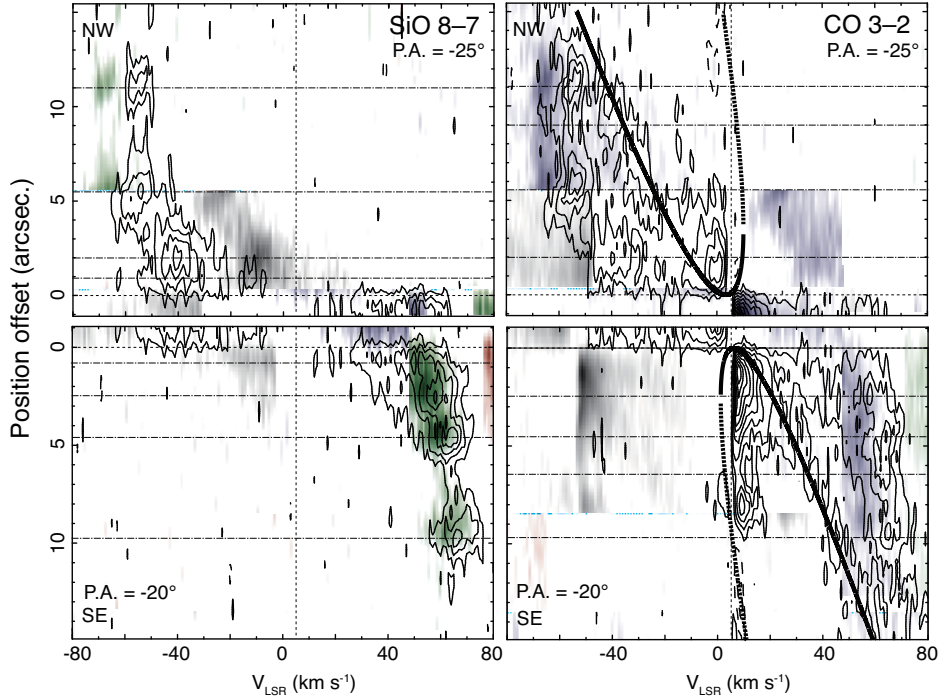}
\caption{Position-velocity (P-V) diagrams of the SiO ({\it left}) and CO ({\it right}) along the jet axes ($-$25$^{\circ}$ in the blueshifted part and $-$20$^{\circ}$ in the redshifted part). Contours are drawn every 3.0 K with the lowest contour at 3.0 K in the SiO, and every 3.5 K with the lowest contour at 3.5 K in the CO. The horizontal thin dashed lines and thin dash-dotted lines are the positions of L1448C(N) and knots, respectively, while the vertical thin dashed lines label the systemic velocity, $V_{\rm sys}=$5.0 km s$^{-1}$. The thick line in the upper right panel denotes the model curve produced by $C=$0.6 arcsec$^{-1}$ and $v_0=$5.0 km s$^{-1}$ arcsec$^{-1}$. The thick line in the lower right panel denote the model curve produced by $C=$0.8 arcsec$^{-1}$ and $v_0=$5.0 km s$^{-1}$ arcsec$^{-1}$. }
\label{fig7}
\end{figure}

\begin{figure}
\plotone{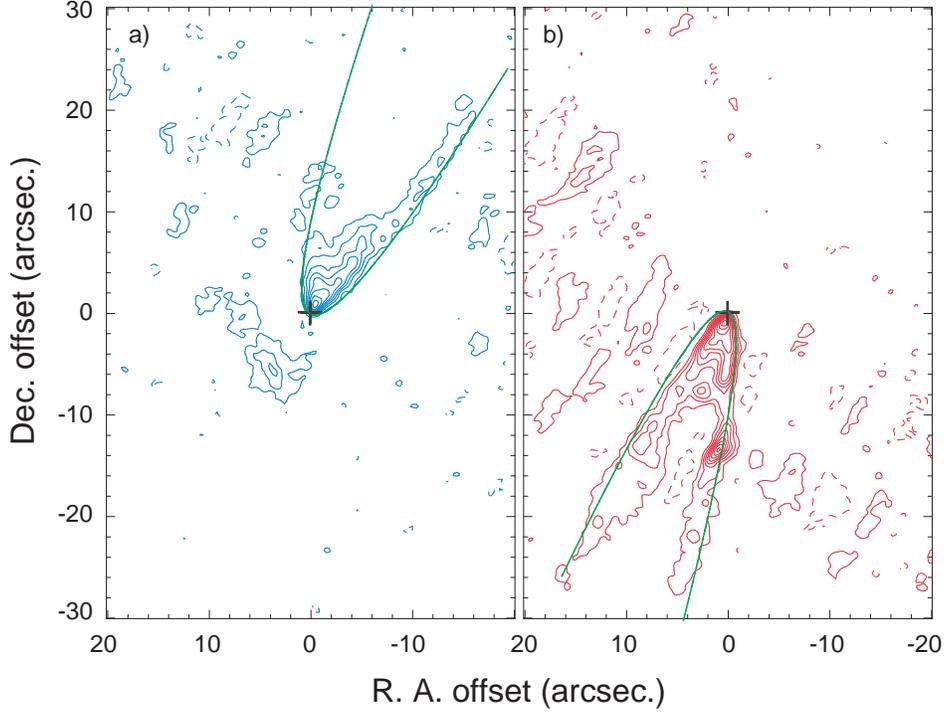}
\caption{Outflow shell model curves overlaid onto the maps of the CO $J$=3--2 emission integrated over the velocity ranges of $\pm$1--40 km s$^{-1}$. 
Contours are drawn every 3.8 Jy beam$^{-1}$ km s$^{-1}$ (4$\sigma$) with the lowest contour of 3.8 Jy beam$^{-1}$ km s$^{-1}$.
a) Blueshifted component with a model curve (green line) with $C=$0.6 arcsec$^{-1}$ and $v_0=$5.0 km s$^{-1}$ arcsec$^{-1}$. 
b) Redshifted component with a model curve (green line) with $C=$0.8 arcsec$^{-1}$ and $v_0=$5.0 km s$^{-1}$ arcsec$^{-1}$.
Both curves were produced by assuming the inclination angle of the axis from the plane of the sky to be 21$^{\circ}$ \citep{Gir01}. }
\label{fig8}
\end{figure}

\begin{figure}
\includegraphics[scale=.65]{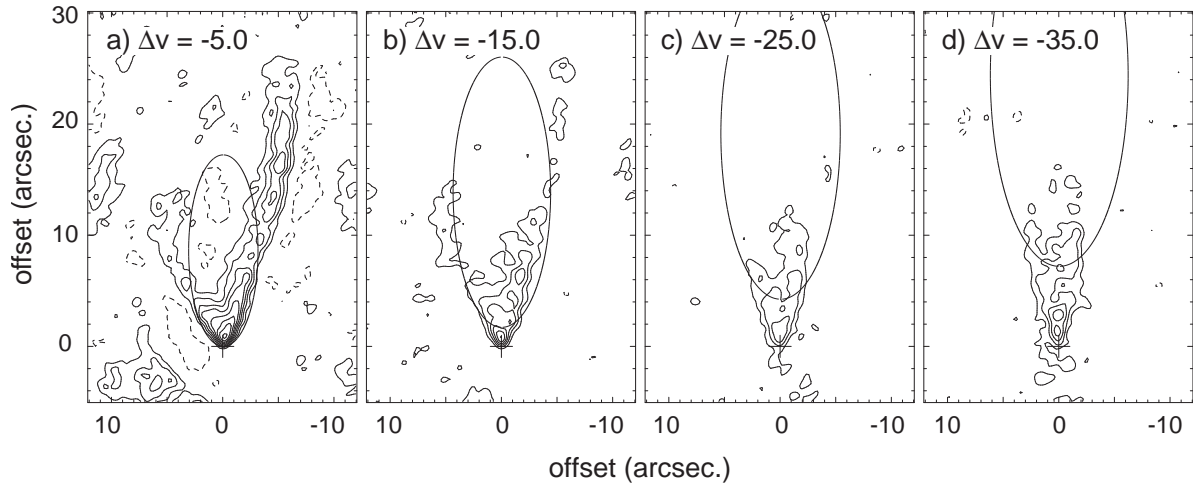}
%\plotone{fig9.eps}
\caption{Outflow shell model curves in the different velocities overlaid onto the maps of the blueshifted CO $J$=3--2 at 10 km s$^{-1}$ wide intervals. The maps are rotated by 25$^{\circ}$ counterclockwise.
The model curve of each velocity was calculated using the same parameters as those of Fig. 8a. }
\label{fig9}
\end{figure}

\begin{figure}
\plotone{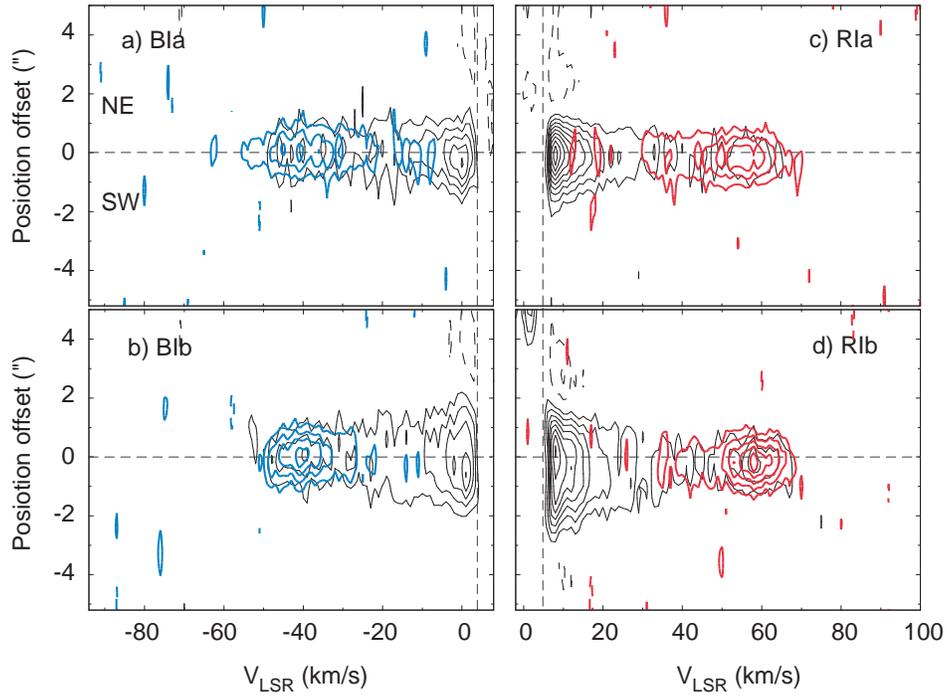}
\caption{Position-velocity (P-V) diagrams of the CO (black contours) and SiO (color contours) along the lines perpendicular to the jet axes. The cuts are across the emission peaks of knots (a) BIa, (b) BIb, c) RIa, and (d) RIb.  Contours are drawn every 3.0 K with the lowest contour at 3.0 K in the SiO, and every 3.5 K with the lowest contour at 3.5 K in the CO. The vertical thin dashed lines label the systemic velocity, $V_{\rm sys}=$5.0 km s$^{-1}$.}
\label{fig10}
\end{figure}

\begin{figure}
\plotone{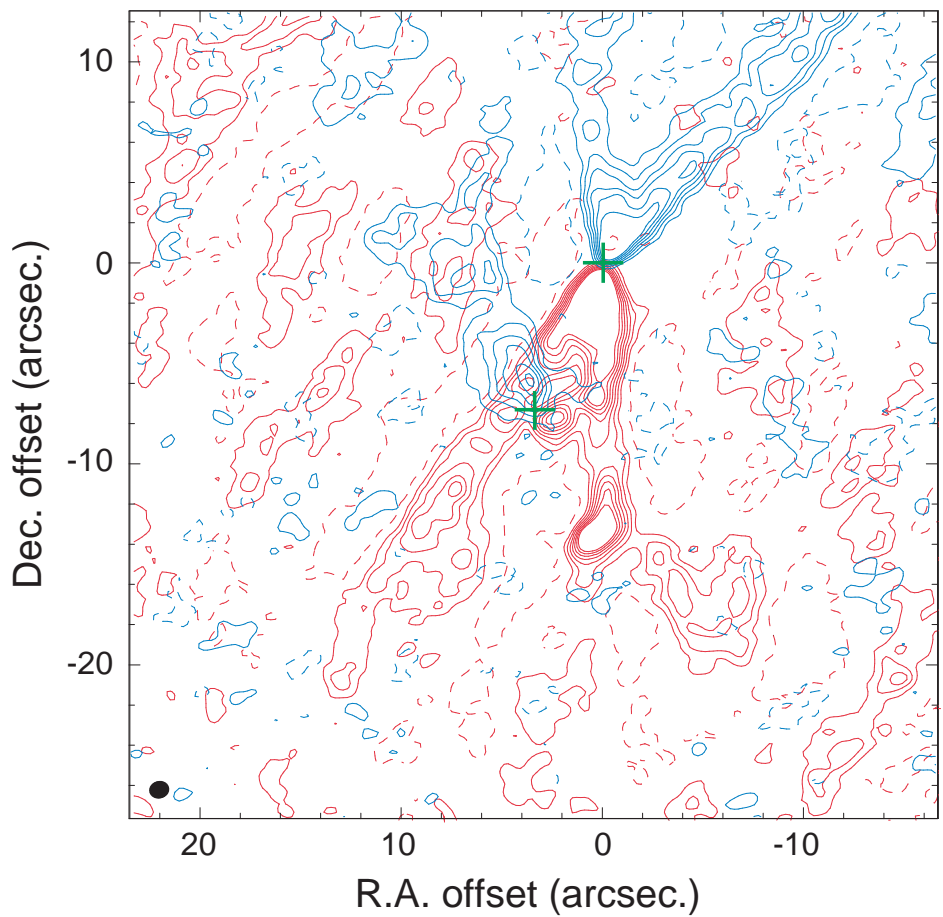}
\caption{Map of the CO $J$=3--2 emission centered at L1448C(S). The velocity offsets from the cloud systemic velocity are from $-$13 to $-$2 km s$^{-1}$ for the blueshiftd component, and from $+$2 to $+$7 km s$^{-1}$ for the redshifted component. Two crosses mark the positions of ptotostars; the southeastern one is L1448C(S) and the northwestern one is L1448C(N). Contours are drawn every 3 $\sigma$ (1.56 Jy beam$^{-1}$ km s$^{-1}$ in the blue and 1.11 Jy beam$^{-1}$ km s$^{-1}$ in the red) intervals with the lowest contours at 3 $\sigma$. Contours for higher flux in the L1448C(N) outflow are eliminated in order to provide the better view of the fainter NE-SW outflow.}
\label{fig11}
\end{figure}

\begin{figure}
\plotone{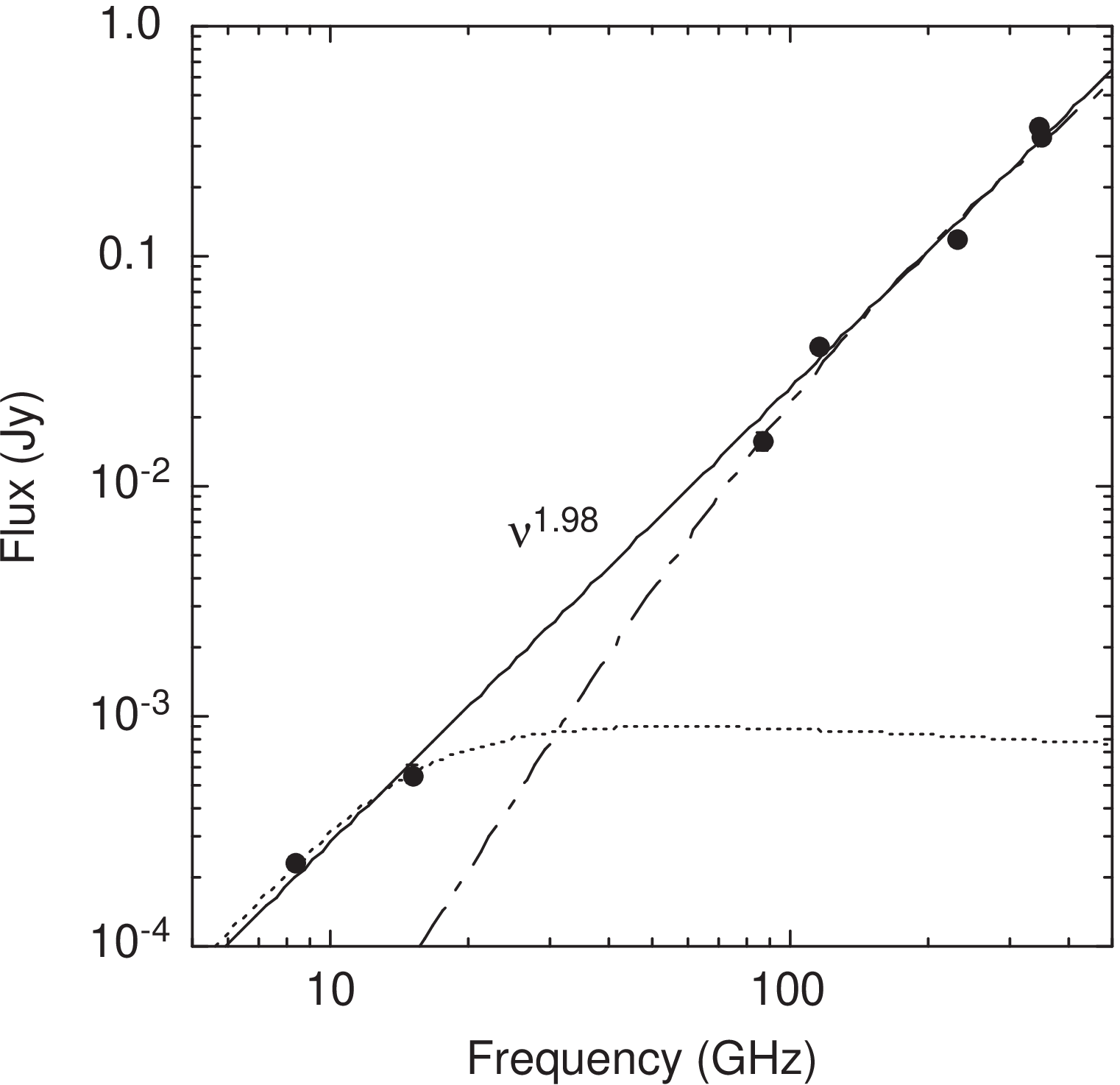}
\caption{SED of the continuum emission from the compact component of L1448C(N).
The filled circles are the observed flux values at 350 GHz (this work),  345 GHz \citep{Jor07}, 230 GHz \citep{Jor07}, 115 GHz \citep{Bac95} , 87 GHz \citep{Gui92}, 15 GHz \citep{Cur90}, and 8.3 GHz \citep{Rei02}. The solid line shows a fit to the observed flux data using $F {\propto}{\nu}^{\alpha}$, where ${\alpha} = 1.98$ is the spectral index. The dotted line is the expected free-free emission calculated using the model of \citet{Cur90}. The dash-dotted curve is the optically thick fit ($S_{\nu}={\Omega}_s B_{\nu}[1-{\rm exp}(-{\tau}_{\nu})]$) for T = 40 K.}
\label{fig12}
\end{figure}

\begin{figure}
\plotone{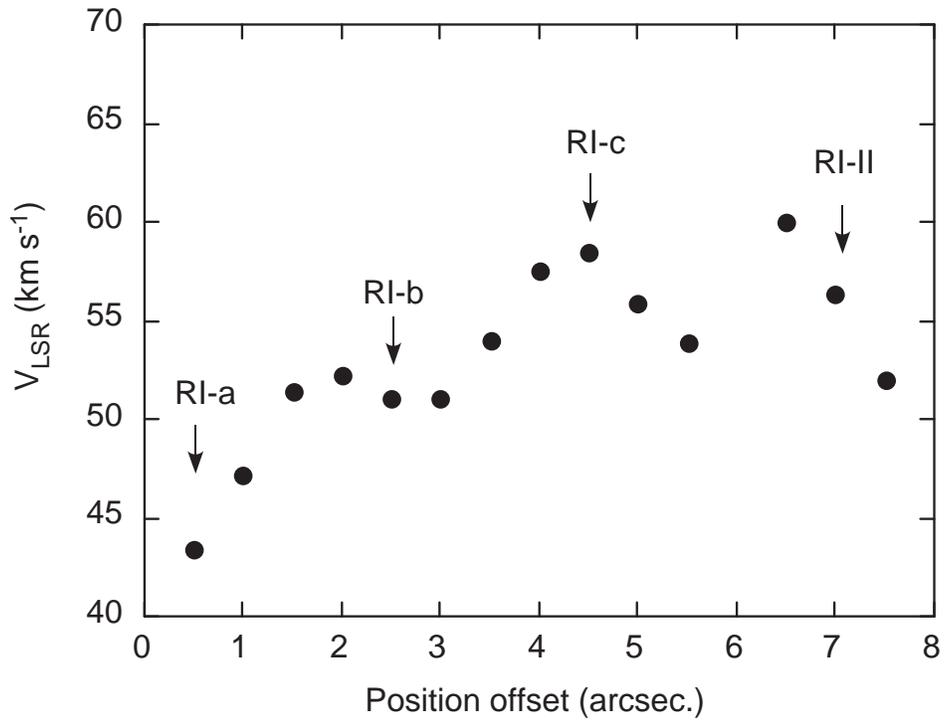}
\caption{Variation of the velocity centroid of the SiO emission in the redshifted part of the jet along the axis (P.A. $-$20$^{\circ}$). }
\label{fig13}
\end{figure}

\end{document}